\documentclass[journal]{IEEEtran}
    \usepackage{balance}
    \usepackage{array}
    \usepackage{booktabs}    \usepackage[caption=false,font=normalsize,labelfont=sf,textfont=sf]{subfig}

    \usepackage[utf8]{inputenc}
    \usepackage[T1]{fontenc}
    \usepackage{lmodern}
    \usepackage{microtype}
    \usepackage{bm}
    \usepackage[hidelinks]{hyperref}

    \usepackage{mathtools, amssymb, amsthm}
    \numberwithin{equation}{section}
    
    \theoremstyle{plain}
    \newif\iffull
\fulltrue

\usepackage{algorithmic, algorithm}
\usepackage[english]{babel}
\usepackage{amssymb}
\usepackage{amsmath,amsfonts}
\usepackage{array}
\usepackage{booktabs}
\usepackage[caption=false,font=normalsize,labelfont=sf,textfont=sf]{subfig}
\usepackage{textcomp}
\usepackage{stfloats}
\usepackage{url}
\usepackage{verbatim}
\usepackage{mathtools}
\usepackage{mathrsfs}
\usepackage{xcolor}
\usepackage{cite}
\usepackage{caption}
\usepackage{subcaption}
\usepackage{amsthm}
\usepackage{bm}
\usepackage{comment}
\usepackage{enumitem}
\usepackage[hidelinks]{hyperref}
\usepackage{thm-restate}

    \newtheorem{theorem}{Theorem}
    \newtheorem{lemma}{Lemma}
    
    \newtheorem{corollary}{Corollary}
    \newtheorem{conjecture}{Conjecture}
    \newtheorem{definition}{Definition}
    \newtheorem{problem}{Problem}
    \newtheorem{example}{Example}
    \newtheorem{remark}{Remark}
    \newtheorem{observation}{Observation}
    
    \usepackage{graphicx}
    \usepackage{tikz}
    \usetikzlibrary{positioning,arrows.meta,decorations.pathreplacing}
    \usepackage{xcolor}
    \usepackage{float}
    \usetikzlibrary{fit,calc}

    \usepackage{hyperref}
    \usepackage{cite}
    
    \newcommand{\E}{\mathbb{E}}

\newcommand{\tight}
{\hspace{-0.25ex}}

\newcommand{\bfv}{{\boldsymbol v}}

\newcommand{\R}{{\Bbb R}}

\usepackage{xcolor}

\usepackage{stackengine}

\def\BibTeX{{\rm B\kern-.05em{\sc i\kern-.025em b}\kern-.08em
    T\kern-.1667em\lower.7ex\hbox{E}\kern-.125emX}}
    
\title{\textbf{General Coverage Models: Structure, Monotonicity, and Shotgun Sequencing}}

\author{Yitzchak Grunbaum,~\IEEEmembership{Student Member,~IEEE} Eitan Yaakobi,~\IEEEmembership{Senior Member,~IEEE}}
    \date{\today}

    \IEEEoverridecommandlockouts

\begin{document}

\maketitle

\begin{abstract}
We study coverage processes in which each draw reveals a subset of $[n]$, and the goal is to determine the expected number of draws until all items are seen at least once. A classical example is the Coupon Collector's Problem, where each draw reveals exactly one item. Motivated by shotgun DNA sequencing, we introduce a model where each draw is a contiguous window of fixed length, in both cyclic and non-cyclic variants. We develop a unifying combinatorial tool that shifts the task of finding coverage time from probability, to a counting problem over families of subsets of $[n]$ that together contain all items, enabling exact calculation. Using this result, we obtain exact expressions for the window models. We then leverage past results on a continuous analogue of the cyclic window model to analyze the asymptotic behavior of both models. We further study what we call uniform $\ell$-regular models, where every draw has size $\ell$ and every item appears in the same number of admissible draws. We compare these to the batch sampling model, in which all $\ell$-subsets are drawn uniformly at random and present upper and lower bounds, which were also obtained independently by Berend and Sher. We conjecture, and prove for special cases, that this model maximizes the coverage time among all uniform $\ell$-regular models. Finally, we prove a universal upper bound on the entire class of uniform $\ell$-regular models, which illuminates the fact that many sampling models share the same leading asymptotic order, while potentially differing significantly in lower-order terms.
\end{abstract}

    \section{Introduction}
    \IEEEPARstart{C}{overage problems} refer to a class of problems in which there are $[n]$ items to draw from, and on each draw a subset is obtained. The question of interest is finding the expected number of draws until all (or some of) the items are collected. One way to describe those problems more rigorously is by considering a hypergraph on $[n]$, with $\{0,1,...,n-1\}$ being the vertices, and all admissible groups we can draw from constituting the hyperedges. Then, the question translates to finding the expected number of draws of hyperedges until all (or a subset of the) vertices are covered, where by being covered we mean that the vertex was contained in one of the drawn hyperedges.

    Perhaps the most basic and well-known example to represent this family of problems is the \emph{Coupon Collector's Problem}.
    The coupon collector's problem, introduced in early probability theory (\tight\cite{demoivre1713, laplace1812}), asks for the expected number of draws needed to collect all $n$ unique coupons under uniform sampling.  In hypergraph language, it means that the hyperedges are singletons, making each $i\in[n]$ have a degree of 1. The solution to this problem is $n H_n$, where $H_n$ is the $n$th harmonic number, which grows as $n \log n$ for large $n$ (\tight\cite{feller1950, bernstein1945}).  Extensions of this problem have considered non-uniform sampling or collecting subsets instead of single elements (\tight\cite{polya, stadje1990, adler2001, flajolet1995}).

Recent works have explored many connections between the coupon collector's problem and DNA storage systems (\tight\cite{coverYourBases, gruica2024, hadas, Roman, Tomer, codingShotgunSequencing}). These include problems such as minimizing coverage depth, ensuring random access, and analyzing fragment overlap structures. In this paper, a connection between \emph{shotgun sequencing} and the coupon collector's problem is drawn. Shotgun sequencing is a powerful method in DNA sequencing, where long DNA molecules are fragmented into overlapping pieces that are sequenced and computationally assembled (\tight\cite{sanger1975, lander2001, venter2001}).

In recent years, shotgun sequencing principles have extended beyond biology into information storage, such as DNA storage systems (\tight\cite{bornholt2016}). These applications motivate probabilistic models that quantify coverage requirements under structured sampling.

A central question in this setting is: how many fragments must be sampled to ensure full coverage of the original sequence? While practical sequencing involves additional considerations such as assembly heuristics and error correction, the probabilistic aspect of achieving complete coverage is a fundamental challenge with implications for both biology and information theory.

Motivated by the structure of shotgun sequencing, we consider a new model in which each observation corresponds to a \emph{consecutive window} within a sequence. Unlike classical models where subsets are arbitrary, consecutive windows introduce strong dependencies that alter the dynamics of coverage. We consider both cyclic and non-cyclic models, and refer to them correspondingly \emph{the cyclic windows coverage problem} and \emph{the windows coverage problem}.

The rest of the paper is organized as follows. Section \ref{sec:defs} introduces the definitions that are used throughout the paper, and states 4 problems, with two of them being the window models mentioned above (Problems \ref{non cyclic windows problem} and \ref{cyclic windows problem}), another problem is a continuous version of Problem \ref{cyclic windows problem} (Problem \ref{continous problem}), and lastly a problem which is referred to as \emph{the batch coverage problem} (Problem \ref{batch sampling problem}) which is the problem of drawing coupons in groups of the same size and any subset is equally likely to be sampled. Inspired by recent work that recasts expectations as combinatorial counts (e.g., \cite{gruica2024}), in Section \ref{anina generalization} we develop a compact, unifying tool that reduces the computation of expected coverage time to counting what we call \emph{recovery sets} in a corresponding hypergraph, which are sets of hyperedges that together contain all vertices.
Section \ref{sec: non cyclic problem} utilizes this tool to solve Problem \ref{non cyclic windows problem}, and presents more compact solutions for special cases. Section \ref{sec:cyclic wondows - solution} is the second to make us of this method, and solves Problem \ref{cyclic windows problem}. Section \ref{cyclic vs continous} makes a comparison between Problems \ref{cyclic windows problem} and \ref{continous problem}, allowing to derive bounds and understand the asymptotic behavior of both Problems \ref{non cyclic windows problem} and \ref{cyclic windows problem}. Section \ref{subsets and monotonicity section} shows lower and upper bounds for the batch coverage model, which were derived essentially at the same time as the independent result of~\cite{berend2025}. It introduces a question regarding all coverage problems for which the corresponding hypergraph is uniform $\ell$-regular, i.e., all hyperedges are of the same size, and all vertices have the same degree, and suggests that a monotonicity property holds: Restricting the drawings to only a subset of the hyperedges, while keeping uniformity might accelerate coverage in expectation, with a proof provided for some cases, and a conjecture that this is always the case and that among all uniform regular models, the batch coverage model achieves the highest expected coverage time. In this sense, the batch model serves as a universal upper bound for the expected time to full coverage across all such models. 
Although a lack of formal proof, partial proofs and empirical evidence strongly supports the conjecture that the batch coverage model indeed forms the upper bound for all uniform regular models. 
Nevertheless, a provable universal upper bound is established, slightly above the batch model’s expectation, guaranteeing that all those coverage models remains within the same asymptotic scale. Lastly, Section \ref{sec:future-work} outlines the remaining questions and a few directions for future research.

    \section{Definitions, Notations And Problem \\Statements}\label{sec:defs}
    For the problems studied in this paper, the \emph{items} to be collected are natural numbers, i.e. for any $n \in \mathbb{N}^+$, define $[n] = \{0,1,\ldots,n-1\}$. These would be considered \emph{collectable} objects, as well as \emph{items, elements} or \emph{positions}. For integers $i < j$, define $[i,j] = \{i, i+1,\ldots, j\}$. For the window models, the following notations are set. For any $n,\ \ell$, define $\mathrm{win}_\ell(i)=[i,i+\ell-1]\subseteq[n]$, a length-$\ell$ \emph{window} starting at $i$, and denote the set of \emph{windows} $\mathcal{W}_{n,\ell}=\{\mathrm{win}_\ell(i):\ i\in [\,n-\ell+1\,]\}$. Similarly,
    for the cyclic case, define $\mathrm{win_{\ell}^\circlearrowright}(i)=\{i,i+1,\dots,i+\ell-1\}\bmod n$, a \emph{cyclic-window} starting at $i$, and define the set of \emph{cyclic-windows} \(\mathcal{W}^{\circlearrowright}_{n,\ell} \coloneqq \{\, \mathrm{win}^{\circlearrowright}_\ell(i) : i\in[n] \,\}
    \). The continuous setting also follows a similar structure. For $\alpha\in[0,1)$ and $a\in(0,1)$, define $\mathrm{arc}_a^{\circlearrowright}(\alpha) = \{\, (\alpha+x)\bmod 1 : x\in[0,a) \,\}$, the length-$a$ \emph{cyclic-arc} starting at $\alpha$, and define the set of cyclic-arcs $\mathcal{A}_a^{\circlearrowright} \coloneqq \{\, \mathrm{arc}_a^{\circlearrowright}(\alpha) : \alpha \in [0,1) \,\}$, and define for any $\gamma\in \mathcal{A}_a^{\circlearrowright}$, its starting point as $\mathrm{Pos}(\gamma)$. Set $H_n := \sum_{k=1}^n \frac{1}{k}$ to be the $n\text{-th}$ harmonic number. For asymptotic results we denote $a_n \approx b_n$ if $\lim_{n \to \infty} \frac{a_n}{b_n} = 1$ and $a_n \lesssim b_n$ if $\displaystyle \limsup_{n\to\infty}\frac{a_n}{b_n}\le 1$ (equivalently, $a_n \le (1+o(1))\,b_n$).

    We are now ready to formally define the problems studied in this paper.     
    \begin{problem}\textbf{The Windows Coverage Problem.}\label{non cyclic windows problem}   
    Consider the process of drawing i.i.d.\ $\omega_1,\omega_2,\dots$ uniformly at random from $\mathcal{W}_{n,\ell}$. The coverage time is the random variable defined as
    \[
    T_{n,\ell} \;=\; \min\Big\{t:\ \bigcup_{r=1}^{t}\omega_r=[n]\Big\}.
    \]
    The problem is to determine the expected coverage time, denoted by \(\E[T_{n,\ell}]\).
    \end{problem}
    
\begin{figure}[t]
    \centering
    \begin{tikzpicture}[x=0.8cm,y=0.9cm,>=latex]
      \def\n{8}    
      \def\ell{3}  
      \def\Wone{3}
      \def\Wtwo{0}
      \def\Wthree{3}
      \def\Wfour{4}
      \def\Wfive{5}
    
      \foreach \x in {0,...,7}{
        \draw[gray!60] (\x,0) -- (\x,-5.8);
        \node[gray!80,below] at (\x,-5.9) {\small \x};
      }
      \draw[gray!80, line width=0.4pt] (-0.3,0.3) -- (7.3,0.3);
      \node[gray!80,above right] at (-0.3,0.3) {\small};
    
      \node[left] at (-0.8,-1.0) {\small $t=1$};
      \node[left] at (-0.8,-2.0) {\small $t=2$};
      \node[left] at (-0.8,-3.0) {\small $t=3$};
      \node[left] at (-0.8,-4.0) {\small $t=4$};
      \node[left] at (-0.8,-5.0) {\small $t=5$};
    
      \newcommand{\drawwin}[3]{%
        \pgfmathtruncatemacro{\winend}{#2+\ell}%
        \draw[fill=purple!20, draw=blue!90, rounded corners=3pt]
          (#2,-#1+0.35) rectangle (\winend,-#1-0.35);
        \node[blue!90] at (#2+1.5,-#1) {\scriptsize $#3$};
      }
    
      \drawwin{1}{\Wone}{[{\Wone},{\Wone+2}]}
      \drawwin{2}{\Wtwo}{[{\Wtwo},{\Wtwo+2}]}
      \drawwin{3}{\Wthree}{[{\Wthree},{\Wthree+2}]}
      \drawwin{4}{\Wfour}{[{\Wfour},{\Wfour+2}]}
      \drawwin{5}{\Wfive}{[{\Wfive},{\Wfive+2}]}
    \end{tikzpicture}
    \caption{Example run of Problem~\ref{non cyclic windows problem} with $n=8$ and $\ell=3$; coverage at $T_{n,\ell}=5$.}
    \label{non cyclic windows figure}
    \end{figure}
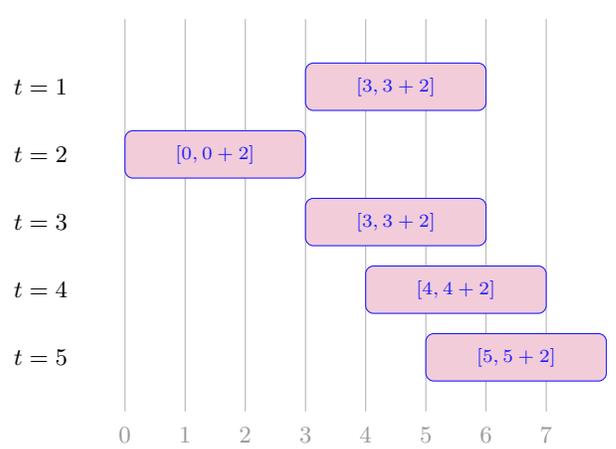

    \begin{example}
    Let $n=8$ and $\ell=3$. 
    Consider the i.i.d.\ draws $\omega_1,\dots,\omega_5 \in \mathcal{W}_{8,3}$ given by their starting positions
    \[
    i_1=3,\quad i_2=0,\quad i_3=3,\quad i_4=4,\quad i_5=5.
    \]
The corresponding windows are 
\(\omega_1=[3,5]\), \(\omega_2=[0,2]\), \(\omega_3=[3,5]\),
\(\omega_4=[4,6]\), and \(\omega_5=[5,7]\).
Their unions after each draw are
\(\{3,4,5\}\),
\(\{0,1,2,3,4,5\}\),
\(\{0,1,2,3,4,5\}\),
\(\{0,1,2,3,4,5,6\}\),
and finally \([n]=\{0,1,2,3,4,5,6,7\}\),
so coverage occurs at time \(T_{n,\ell}=5\). An illustration appears in Figure \ref{non cyclic windows figure}. 
    \end{example}
    
    The cyclic version of Problem~\ref{non cyclic windows problem} is defined next.    
    \begin{problem}\label{cyclic windows problem}
    \textbf{The Cyclic Windows Coverage Problem.} Consider the process of drawing i.i.d.\ $\nu_1,\nu_2,\dots$ uniformly at random from $\mathcal{W}_{n,\ell}^\circlearrowright$. The cyclic coverage time is the random variable defined as
    \[
    T_{n,\ell}^\circlearrowright
     \;=\; \min\Big\{t:\ \bigcup_{r=1}^{t}\nu_r= [n]\Big\}.
    \]
    The problem is to determine the expected cyclic coverage time, denoted by \(\E[T_{n,\ell}^\circlearrowright]\).
    \end{problem}
    
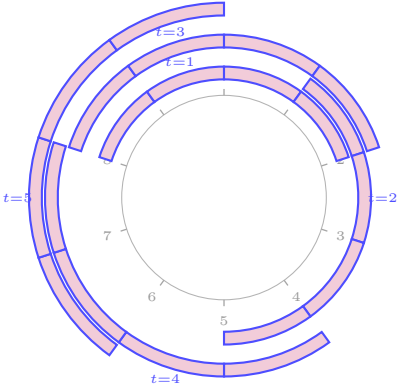
\begin{figure}[t]
\centering
\begin{tikzpicture}[scale=0.85]

  \def\n{10}\def\ell{4}\def\R{1.6}
  \colorlet{winfill}{purple!20}
  \colorlet{winborder}{blue!70}
  \tikzset{winstyle/.style={line width=0.8pt, draw=winborder, fill=winfill}}

  \def\Vone{8}\def\Vtwo{1}\def\Vthree{8}\def\Vfour{4}\def\Vfive{6}

  \draw[gray!60, line width=0.4pt] (0,0) circle (\R);
  \foreach \i in {0,...,9}{
    \pgfmathsetmacro{\ang}{90 - \i*360/\n}
    \draw[gray!70, line width=0.5pt] (\ang:\R) -- (\ang:\R+0.1);
    \node[gray!80, font=\tiny] at (\ang:\R+0.32) {\i};
  }

  \newcommand{\winband}[2]{%
    \pgfmathsetmacro{\radin}{\R + 0.25*(#1)}       
    \pgfmathsetmacro{\radout}{\radin + 0.20}        
    \foreach \k in {0,...,\numexpr\ell-1} {
      \pgfmathsetmacro{\idx}{mod(#2 + \k, \n)}
      \pgfmathsetmacro{\start}{90 - \idx*360/\n}
      \pgfmathsetmacro{\end}{\start - 360/\n}
      \draw[winstyle] (\start:\radin)
        arc[start angle=\start, end angle=\end, radius=\radin] --
        (\end:\radout)
        arc[start angle=\end, end angle=\start, radius=\radout] -- cycle;
    }
    \pgfmathsetmacro{\mididx}{mod(#2 + 0.5*(\ell-1), \n)}
    \pgfmathsetmacro{\midang}{90 - \mididx*360/\n}
    \node[winborder, font=\tiny] at (\midang:\radout+0.18) {$t{=}#1$};
  }

  \winband{1}{\Vone}
  \winband{2}{\Vtwo}
  \winband{3}{\Vthree}
  \winband{4}{\Vfour}
  \winband{5}{\Vfive}

\end{tikzpicture}
\caption{Example run of Problem~\ref{cyclic windows problem} with $n=10$ and $\ell=4$; coverage at $T_{n,\ell}^\circlearrowright=4$.}
\label{cyclic windows figure}
\end{figure}

    \begin{example}
    Let $n=10$ and $\ell=4$. 
    Consider the i.i.d.\ draws $\nu_1,\dots,\nu_5\in\mathcal{W}^{\circlearrowright}_{10,4}$ given by their starting positions
    \[
    i_1=8,\quad i_2=1,\quad i_3=8,\quad i_4=4,\quad i_5=6,
    \]
    so that
\(\nu_1=\{8,9,0,1\}\), \(\nu_2=\{1,2,3,4\}\), \(\nu_3=\{8,9,0,1\}\),
\(\nu_4=\{4,5,6,7\}\), and \(\nu_5=\{6,7,8,9\}\). Their unions after each draw are
  \(\{8,9,0,1\}\),
\(\{0,1,2,3,4,8,9\}\),
\(\{0,1,2,3,4,8,9\}\),
and finally \([n]=\{0,1,2,3,4,5,6,7,8,9\}\),
 and hence coverage occurs at time $T_{n,\ell}^{\circlearrowright}=4$ (the fifth draw is redundant).
    An illustration appears in Figure~\ref{cyclic windows figure}.
    \end{example}

    A continuous version of Problem~\ref{cyclic windows problem} is defined as follows.     
    \begin{problem} \label{continous problem}
    \textbf{The Arcs Coverage Problem.} Fix a circle with a circumference of length 1. Fix an arc length $a\in(0,1)$. Consider the process of drawing i.i.d.\ $\gamma_1,\gamma_2,\dots$ uniformly at random from $\mathcal{A}_a^\circlearrowright$, and put each $\gamma_i$ on the circle's circumference starting at point $\mathrm{Pos}(\gamma_i)$ and going clockwise until the entire circumference, which can be described as $[0,1)$ mod 1 cyclically, is covered. The coverage time is the random variable which is defined as
    \[
    \mathcal{T}_a \;=\; \min\Big\{t:\ \bigcup_{r=1}^{t} \gamma_r =[0,1)\Big\}.
    \]
    The problem is to determine the expected coverage time, denoted by \(\E[\mathcal T_a]\).
    \end{problem}
    
    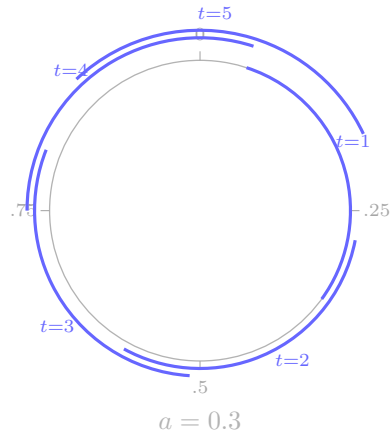
\begin{figure}[!h]
    \centering
    \begin{tikzpicture}[scale=1.0]
      \def\R{2.0}   
      \def\a{0.3}   
    
      \def\Uone{0.05}
      \def\Utwo{0.28}
      \def\Uthree{0.51}
      \def\Ufour{0.75}
      \def\Ufive{0.88}
    
      \draw[gray!60, line width=0.5pt] (0,0) circle (\R);
    
      \foreach \u/\lab in {0/{0},0.25/{.25},0.5/{.5},0.75/{.75}}{
        \pgfmathsetmacro{\ang}{90 - 360*\u}
        \draw[gray!70] (\ang:\R) -- (\ang:\R+0.12);
        \node[gray!70, font=\scriptsize] at (\ang:\R+0.35) {\lab};
      }
    
      \newcommand{\drawArcSameR}[3]{
        \pgfmathsetmacro{\startang}{90 - 360*(#2)}
        \pgfmathsetmacro{\deltaang}{-360*\a}  
        \draw[line width=1.2pt, draw=#3]
          (\startang:\R) arc[start angle=\startang, delta angle=\deltaang, radius=\R];
        \pgfmathsetmacro{\midang}{\startang + 0.5*\deltaang}
        \node[font=\scriptsize, #3] at (\midang:\R+0.22) {$t{=}#1$};
      }
    
      TINY OFFSET version (uncomment next macro + calls if you prefer slight stacking)
      \newcommand{\drawArcTinyOffset}[3]{
        \pgfmathsetmacro{\startang}{90 - 360*(#2)}
        \pgfmathsetmacro{\deltaang}{-360*\a}
        \pgfmathsetmacro{\rad}{\R + 0.10*(#1-1)}  
        \draw[line width=1.2pt, draw=#3]
          (\startang:\rad) arc[start angle=\startang, delta angle=\deltaang, radius=\rad];
        \pgfmathsetmacro{\midang}{\startang + 0.44*\deltaang}
        \node[font=\scriptsize, #3] at (\midang:\rad+0.24) {$t{=}#1$};
      }
    
    
      If you prefer the tiny-offset look, comment the 5 lines above and use:
      \drawArcTinyOffset{1}{\Uone}{blue!60}
      \drawArcTinyOffset{2}{\Utwo}{blue!60}
      \drawArcTinyOffset{3}{\Uthree}{blue!60}
      \drawArcTinyOffset{4}{\Ufour}{blue!60}
      \drawArcTinyOffset{5}{\Ufive}{blue!60}
    
      \node[gray!60] at (0,-\R-0.8) {$a=0.3$};
    \end{tikzpicture}
    \caption{Example run of Problem~\ref{continous problem} with $a=0.3$. On round $t$, the arc $I_t=[U_t,U_t+a)\pmod{1}$ is placed; coverage at $\mathcal{T}_{0.3}=5$.}
    \label{continous figure}
    \end{figure}
    
    \begin{example}
    Fix $a=0.3$. Consider i.i.d.\ draws $\gamma_1,\dots,\gamma_4\in\mathcal{A}_a^{\circlearrowright}$ with starting points
    \[
    x_1=0.05,\quad x_2=0.28,\quad x_3=0.51,\quad x_4=0.75,
    \]
   so that, with $\gamma_t=\mathrm{arc}_a^{\circlearrowright}(x_t)=[x_t,x_t+a)\pmod{1}$,  
the arcs are 
\(\gamma_1=[0.05,0.35)\), \(\gamma_2=[0.28,0.58)\),
\(\gamma_3=[0.51,0.81)\), and \(\gamma_4=[0.75,1)\cup[0,0.05)\).
The unions after each draw are
\([0.05,0.35)\), \([0.05,0.58)\),
\([0.05,0.81)\), and finally \([0,1)\).
 Hence coverage occurs at time $\mathcal{T}_{0.3}=4$.
    An illustration appears in Figure \ref{continous figure}.
    \end{example}
    The continuous model was introduced and analyzed in~\cite{stevens1939,flatto1962,after flatto - Steutel1967,solomon1978},
    where the solution for this problem was introduced by
    \[
    \E[\mathcal T_a]= 1+\sum_{k=1}^{\lfloor 1/a \rfloor}
    (-1)^{k-1}\,\frac{(1-ka)^{\,k-1}}{(ka)^{\,k+1}},
    \]
     and in~\cite{flatto1962} and \cite{after flatto - Steutel1967}, it was shown that \[\E[\mathcal{T}_a]=\tfrac{1}{a}\big( \,\log\!\tfrac{1}{a} +\log\log\tfrac{1}{a}+\gamma + o(1)\big)
    \text{, as } a\rightarrow 0.\]
    Section~\ref{cyclic vs continous} establishes its relationship to Problem~\ref{cyclic windows problem}, as this problem presents a model that can be considered as the continuous version of the cyclic windows coverage problem. The last problem we consider in this paper is defined next.
    \begin{problem}\label{batch sampling problem}
    \textbf{The Batch Coverage Problem.}
    Fix $n$ and $\ell$. Consider the process of drawing subsets $S_1,S_2,\ldots$, referred to as batches, of size $|S_i|=\ell$ uniformly from $\{S\subseteq[n]:|S|=\ell\}$.
    The coverage time is defined as
    \[
    T_{\binom{[n]}{\ell}} \;=\; \min\Big\{t:\ \bigcup_{r=1}^{t}S_r=[n]\Big\}.
    \]
The problem is to determine the expected coverage time, denoted by \(\E[T_{\binom{[n]}{\ell}}]\).

    \end{problem}

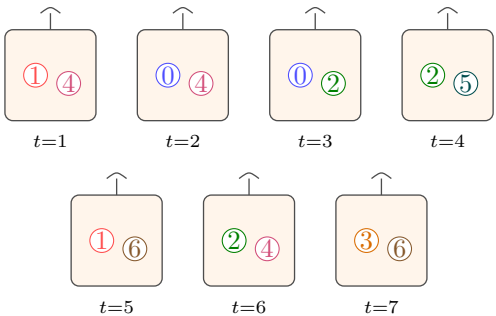
\begin{figure}[b]
    \centering
    \begin{tikzpicture}[x=1.1cm,y=1.1cm,>=Latex]
    
      \newcommand{\numcolorname}[1]{%
        \ifcase#1 blue!70%
        \or red!70%
        \or green!50!black%
        \or orange!85!black%
        \or purple!70%
        \or teal!60!black%
        \or brown!70!black%
        \else gray!70\fi}
    
      \newcommand{\candybagxy}[5]{%
        \begin{scope}[shift={(#1,#2)}]
          \draw[rounded corners=3pt, draw=black!70, fill=orange!8, line width=0.5pt]
            (-0.55,-0.55) rectangle (0.55,0.55);
          \draw[black!70, line width=0.5pt] (0,0.55) -- (0,0.75);
          \draw[black!70, line width=0.5pt] (-0.12,0.75) .. controls (0,0.84) .. (0.12,0.75);
          \node[circle, draw=\numcolorname{#4}, fill=white, minimum size=9pt, inner sep=0pt]
               at (-0.18,0.00) {\textcolor{\numcolorname{#4}}{#4}};
          \node[circle, draw=\numcolorname{#5}, fill=white, minimum size=9pt, inner sep=0pt]
               at ( 0.22,-0.10) {\textcolor{\numcolorname{#5}}{#5}};
          \node[font=\scriptsize, anchor=north] at (0,-0.62) {$t{=}$#3};
        \end{scope}
      }
    
      \def\dx{1.6}
      \def\dy{-2.0}
    
      \candybagxy{0*\dx}{0}{1}{1}{4}
      \candybagxy{1*\dx}{0}{2}{0}{4}
      \candybagxy{2*\dx}{0}{3}{0}{2}
      \candybagxy{3*\dx}{0}{4}{2}{5}
    
      \candybagxy{0.5*\dx}{\dy}{5}{1}{6}
      \candybagxy{1.5*\dx}{\dy}{6}{2}{4}
      \candybagxy{2.5*\dx}{\dy}{7}{3}{6}
    
    \end{tikzpicture}
    \caption{Example run of \ref{batch sampling problem} with $n=7$ and $\ell=2$; coverage at $T_{\binom{[n]}{\ell}}=7$.}
    \label{batch sampling figure}
    \end{figure}

    \begin{example}
    Let $n=7$ and $\ell=2$. 
    Consider the i.i.d.\ draws $S_1,\dots,S_7\subseteq[n]$ with $|S_t|=2$ given by
    \(S_1=\{1,4\}\), \(S_2=\{0,4\}\), \(S_3=\{0,2\}\), \(S_4=\{2,5\}\), \(S_5=\{1,6\}\), \(S_6=\{2,4\}\) and \(S_7=\{3,6\}
    \). Their unions after each draw are
    \(\{1,4\}\), \(\{0,1,4\}\), \(\{0,1,2,4\}\),
    \(\{0,1,2,4,5\}\), \(\{0,1,2,4,5,6\}\),
     \(\{0,1,2,4,5,6\}\) and \(\{0,1,2,3,4,5,6\}=[n].
    \)
    Thus coverage occurs at time $T_{\binom{[n]}{\ell}}=7$. An illustration appears in Figure \ref{batch sampling figure}.
    \end{example}

    This model has been studied in \cite{stadje1990,adler2001, polya}, where combinatorial solutions for the moments of $T_{\binom{[n]}{\ell}}$ and particularly for the expectation are given. The first solution for the expectation was given by P\'olya in \cite{polya}, where he proved that the expectation is
    \[
    \E[T_{\binom{[n]}{\ell}}]=\binom{n}{\ell} \sum_{i=0}^{n-1} \frac{(-1)^{n-i+1} \binom{n}{i}}{\binom{n}{\ell} - \binom{i}{\ell}},
    \]
    and analyzed asymptotics for the case \(\ell=O(1)\), and obtained for this case that
    \[
    \E[T_{\binom{[n]}{\ell}}]\approx \big( \tfrac{n+\frac{1}{2}}{\ell}-\tfrac{1}{2} \big)\big( \log n + \gamma \big) + \tfrac{1}{2} +o(1).
    \]
    A more general setting was solved by \cite{stadje1990}. Given
    \(A\subseteq[n]\) with \(|A|=m\), let \(Z_k(A)\) be the number of draws
    needed to see at least \(k\le m\) distinct elements of \(A\).
    The exact formula was obtained
    \[
\mathbb{E}[Z_k(A)]
\tight=\tight\binom{n}{\ell}\tight\tight\sum_{i\tight=\tight0}^{k\tight-\tight1}
\tight
\tight\frac{(-1)^{\tight k\tight-\tight i\tight+\tight1}\tight\binom{m}{i}\tight\binom{m\tight-\tight i\tight-\tight1}{m\tight-\tight k}}{\binom{n}{\ell}\tight-\tight\binom{i\tight+\tight n\tight-\tight m\tight}{\ell}}.
\] Section~\ref{subsets and monotonicity section} presents a more transparent expression valid for all $1\le\ell\le n$, and discusses its relationship to all other models of coverage.

    \section{A General Combinatorial Approach for \\Solving Coverage Problems}\label{anina generalization}
    Before we study the proposed coverage problems in this paper, we first seek to establish a general combinatorial approach which will be used in our analysis throughout the paper. This approach, motivated by the work of~\cite{gruica2024}, produces a tool which converts the problem of calculating the expectation of the coverage time to a combinatorial problem in which the expectation can be expressed in a combinatorial fashion. While the motivation in~\cite{gruica2024} was to study the so-called \emph{random access problem} in DNA-based storage systems, the approach can be extended in a more general form.

    All coverage problems share a common structure in which the expected time to full coverage admits a single, model-independent expression that only depends on the number of \emph{recovery sets} that will be defined next. Thus, the task of computing the expectation reduces to a combinatorial problem of counting recovery sets. First, we show a general framework to present coverage problems using hypergraphs. 

\begin{definition}
    A \emph{hypergraph} is a pair $\mathcal{H} = (V,E)$ where
    \begin{itemize}
        \item $V$ is a finite set of vertices, and
        \item $E \subseteq 2^V$ consists of non-empty subsets of $V$, called \emph{hyperedges}.
    \end{itemize}
    A hypergraph is said to be \emph{covering} if $\bigcup_{e \in E} e = V$.

\end{definition}
    
    \begin{definition}
    Let $\Omega$ be a set of $n$ coverable objects, i.e., $|\Omega|=n$. A coverage model is specified
    by a finite family of \emph{observable blocks} $\mathcal{A}=\{A_1,\dots,A_{N_\mathcal{H}}\}$, where each
    $A_i\subseteq\Omega$ is the set of objects revealed by a single draw.
    Equivalently, the coverage model can be presented using a hypergraph
    \[
    \mathcal{H}=(V,E),\qquad V=\Omega,\quad E=\mathcal{A}\subseteq 2^{V},
    \]
    where the vertices are the objects and the hyperedges are the observable blocks. The hypergraph $\mathcal{H}$ is assumed to be \emph{covering}. On each round, an index $I_r\in[N_\mathcal{H}]$ is uniformly and independently sampled, revealing the $I_r$-th hyperedge and \emph{collecting} its vertices $A_{I_r}$. The coverage time is defined to be
    \[
    T_{\mathcal{H}} \;=\; \min\Big\{t:\ \bigcup_{r=1}^{t}A_{I_r} \,=\, \Omega\Big\}.
    \]
    A \emph{recovery set} is a subset of the hyperedges $\mathcal{R}\subseteq \mathcal{A}$ such that \(\bigcup_{\mathcal{E}\in\mathcal{R}}\mathcal{E}=\Omega\). The set of all recovery sets is defined as \(
    \mathrm{Rec}(\mathcal{A})
    := \big\{\, \mathcal{R}\subseteq \mathcal{A}\ :\mathcal{R} \text{ is a recovery set}\}.
    \)
    Denote \(M_\mathcal{H}\) as the maximal size of any non-recovery set, that is,
    $M_\mathcal{H}= \max\Big\{\, |\mathcal{R}|\ :\mathcal{R}\in2^\mathcal{A}\setminus \mathrm{Rec}(\mathcal{A})\}$. And denote $\alpha_\mathcal{H}(s)$ to be the number of $s$-sized recovery sets. Explicitly 
    $\alpha_\mathcal{H}(s)=|\{\mathcal{R}\in \mathrm{Rec}(\mathcal{A}):|\mathcal{R}|=s\}|$.
    When the hypergraph is obvious from context, we omit the \(_\mathcal{H}\) subscript.
    \end{definition}
    The following examples illustrate how a coverage problem can be presented by a hypergraph.

    \begin{example}\label{coverage example classical}
      {Classical Coupon Collector's Problem.}
      \(V=[n]\), \(E=\{\{i\}: i\in V\}\), \(N_\mathcal{H}=n\).
      Each draw reveals a singleton; coverage means every vertex has been sampled.
    \end{example}
    For binary vectors $u,v \in \{0,1\}^d$, their \emph{Hamming distance} is \(\operatorname{dist}_H(u,v) = |\{\,i \in [d] : u_i \ne v_i\,\}|.\) For a binary vector $v \in \{0,1\}^d$ and radius $t \in \mathbb{N}$, the \emph{Hamming ball} of radius $t$ centered at $v$ is \(B_t(v) = \{\,u \in \{0,1\}^d : \operatorname{dist}_H(u,v) \le t\,\}.\)

\begin{example} \label{coverage example hamming}
      {Hamming Hypercube Coverage.}
Fix $d,t \in \mathbb{N}$.  
Let \(V = \{0,1\}^d\) be the set of vertices of the $d$-dimensional Hamming hypercube, so $|V| = 2^d$.  
Let
\[
E = \{\,B_t(x) : x \in V\,\}
\]
be the family of all Hamming balls of radius \(t\).  
Then \(N_\mathcal{H} = |V| = 2^d\).  
Each draw reveals a Hamming ball of radius \(t\); coverage means that every vertex lies in some revealed ball.
\end{example}

    \begin{theorem}\label{thm:hypergraph-expectation}
    Given a hypergraph \(\mathcal{H}\),
    \[
    \mathbb{E}[T_{\mathcal{H}}]= N_\mathcal{H}\big(H_{N_\mathcal{H}}-H_{N_\mathcal{H}-M_\mathcal{H}-1}\big)-\sum_{s=0}^{M_\mathcal{H}}\frac{\alpha_\mathcal{H}(s)}{\binom{N_\mathcal{H}-1}{s}}.
    \]
    \end{theorem}

    \begin{IEEEproof}[Proof (following \cite{gruica2024})]
    By the tail-sum formula,
    \(
    \E[T_{\mathcal{H}}] \;=\; \sum_{r\ge 1}\Pr\!\big(T_{\mathcal{H}}\ge r\big).
    \)
    Let $\mu_r$ denote the number of distinct hyperedges revealed after $r$ rounds. By the law of total probability and since any set of size greater than $M_\mathcal{H}$ is a recovery set, \(
    \Pr\!\big(T_{\mathcal{H}}\ge r\big)
    = \sum_{s=0}^{M_\mathcal{H}} \Pr\!\big(T_{\mathcal{H}}\ge r \mid \mu_{r-1}=s\big)\,\Pr\!\big(\mu_{r-1}=s\big).
    \)
    Conditioned on $\mu_{r-1}=s$,
    \(
    \Pr\!\big(T_{\mathcal{H}}\ge r \mid \mu_{r-1}=s\big)
    = 1 - \frac{\alpha_\mathcal{H}(s)}{\binom{N_\mathcal{H}}{s}},
    \)
    since $\alpha_\mathcal{H}(s)$ counts the size-$s$ recovery sets among the $\binom{N_\mathcal{H}}{s}$ many $s$-subsets. Moreover, by the inclusion–exclusion principle,
    \[
    \Pr\!\big(\mu_{r-1}=s\big)
    = \binom{N_\mathcal{H}}{s}\sum_{j=0}^{s}\binom{s}{j}(-1)^j\Big(\frac{s-j}{N_\mathcal{H}}\Big)^{r-1}.
    \]
    Therefore,
    \begin{align*}
    \E[T_{\mathcal{H}}]
    &= \sum_{r\ge 1}\tight\tight\sum_{s=0}^{M_\mathcal{H}}\tight\tight\Big(\tight1\tight-\tight\frac{\alpha_\mathcal{H}(s)}{\binom{N_\mathcal{H}}{s}}\tight\tight\Big)\tight\binom{N_\mathcal{H}}{s}\tight\tight\tight
    \sum_{j=0}^{s}\tight\tight\binom{s}{j}\tight\tight(-1)^j\tight\tight\Big(\tight\frac{s-j}{N_\mathcal{H}}\tight\Big)^{\tight\tight\tight r-1}\\
    &= \sum_{s=0}^{M_\mathcal{H}}\big(\binom{N_\mathcal{H}}{s}-\alpha_\mathcal{H}(s)\big)
    \sum_{j=0}^{s}\binom{s}{j}(-1)^j \sum_{r\ge 0}\Big(\frac{s-j}{N_\mathcal{H}}\Big)^{r}\\
    &= \sum_{s=0}^{M_\mathcal{H}}\big(\binom{N_\mathcal{H}}{s}-\alpha_\mathcal{H}(s)\big)
    \sum_{j=0}^{s}\binom{s}{j}(-1)^j \frac{N_\mathcal{H}}{\,N_\mathcal{H}-s+j\,}.
    \end{align*}
    Using the identity $\sum_{j=0}^{s}\binom{s}{j}(-1)^j \frac{N_\mathcal{H}}{\,N_\mathcal{H}-s+j\,}
    = \frac{1}{\binom{N_\mathcal{H}-1}{s}}$, we get
    \begin{align*}
    \E[T_{\mathcal{H}}]
    &= \sum_{s=0}^{M_\mathcal{H}}\frac{\binom{N_\mathcal{H}}{s}-\alpha_\mathcal{H}(s)}{\binom{N_\mathcal{H}-1}{s}}
    = \sum_{s=0}^{M_\mathcal{H}}\frac{\binom{N_\mathcal{H}}{s}}{\binom{N_\mathcal{H}-1}{s}}
    - \sum_{s=0}^{M_\mathcal{H}}\frac{\alpha_\mathcal{H}(s)}{\binom{N_\mathcal{H}-1}{s}}
    \\&= N_\mathcal{H}\sum_{s=0}^{M_\mathcal{H}}\frac{1}{N_\mathcal{H}-s}
    - \sum_{s=0}^{M_\mathcal{H}}\frac{\alpha_\mathcal{H}(s)}{\binom{N_\mathcal{H}-1}{s}}
    \\&= N_\mathcal{H}\big(H_{N_\mathcal{H}}-H_{N_\mathcal{H}-M_\mathcal{H}-1}\big)
    - \sum_{s=0}^{M_\mathcal{H}}\frac{\alpha_\mathcal{H}(s)}{\binom{N_\mathcal{H}-1}{s}}.
    \end{align*}
    \end{IEEEproof}
    This approach reduces the task of determining the expected coverage time to computing \(M\) and \(\alpha_\mathcal{H}(s)\).
    Once these are known, the expectation follows immediately from Theorem~\ref{thm:hypergraph-expectation},
    turning the probabilistic question into a purely combinatorial calculation.
    
    Theorem \ref{thm:hypergraph-expectation} is now used to compute the expectation in Examples \ref{coverage example classical} and \ref{coverage example hamming}, illustrating how straightforward this approach is in practice.

    \vspace{1ex}
\noindent\textbf{Example~\ref{coverage example classical} (continued).}
\emph{Classical Coupon Collector's Problem. There are no recovery sets of size $s\le n-1$, i.e. $\alpha(s)=0$ for $s\le n-1$. Hence, by Theorem \ref{thm:hypergraph-expectation}, the expected number of draws is $nH_n$.
}
\vspace{0.5\baselineskip}

\noindent\textbf{Example~\ref{coverage example hamming} (continued).}
\emph{Hamming Hypercube Coverage.
    The maximum size of a
    non–recovery set is
    \(M \;=\; N-|B_t|\;=\;2^d-\sum_{j=0}^{t}\binom{d}{j},\)
    since a vertex $v$ remains uncovered if and only if all balls containing $v$ are omitted.
    Hence, by Theorem~\ref{thm:hypergraph-expectation}, the expected number of draws is
    \[
    2^d\big(H_{2^d}-H_{\,|B_t|-1}\big)
    \;-\;
    \sum_{s=0}^{\,2^d-|B_t|}\frac{\alpha(s)}{\binom{2^d-1}{s}},
    \]
    where $\alpha(s)$ counts all size-$s$ collections of radius-$t$ balls whose union is the entire cube.}
    
    \vspace{0.5\baselineskip}
    
    From here on, for every coverage model throughout the paper, we will refer to the possible attainable subsets of collectable items as the hyperedges of the graph of items interchangeably.
\section{Problem~\ref{non cyclic windows problem}: The Windows Coverage Model} \label{sec: non cyclic problem}
    In this section we solve Problem~\ref{non cyclic windows problem}, the window coverage problem.

    \subsection{General Solution to The Windows Coverage Problem}
    Using the approach derived in Section~\ref{anina generalization}, Problem~\ref{non cyclic windows problem} can be formulated using hypergraphs as follows. Denote
    \[
    \mathcal{H}_{n,\ell}=(V,E),\qquad
    V=[n],\qquad
    E=\mathcal{A}=\mathcal{W}_{n,\ell},
    \]
    so it holds that $N=|E|=n$ and $M=n-\ell$ (omitting the coverage of the rightmost position admits the largest non recovery set). Let \(\alpha(s)\) denote the number of size-\(s\) recovery sets in \(\mathcal{W}_{n,\ell}\). Then, by Theorem~\ref{thm:hypergraph-expectation} it holds that,
    \begin{equation}\label{eq:ET-window}
    \;
    \E[T_{n,\ell}]
    = (n-\ell+1)\,H_{\,n-\ell+1}
    \;-\;
    \sum_{s=0}^{\,n-\ell}\frac{\alpha(s)}{\binom{\,n-\ell\,}{s}}.
    \;
    \end{equation}
    The task then reduces to determining the values of \(\alpha(s)\), yielding the following.
    \begin{lemma}\label{lem1}

    It holds that
    \begin{align*}
\mathbb{E}[T_{n,\ell}] =& (n - \ell + 1) H_{n - \ell + 1} \\
& - \tight\tight\sum_{s = \left\lceil\tight\tight \frac{n}{\ell} \right  \rceil}^{n - \ell}
    \tight\tight\tight\sum_{i=0}^{\left\lfloor \frac{n - \ell - s + 1}{\ell} \right\rfloor}
    \tight\tight\tight\tight\tight\tight\tight(-1)^i \binom{s - 1}{i}
    \tight\cdot\tight \frac{ \binom{n - \ell - i\ell - 1}{s - 2}}{\binom{n - \ell}{s}}.
\\
\text{In particular,} \\
\mathbb{E}[T_{n,2}] =& (n-1)\,H_{\,n-1}
- \sum_{s=\lceil \frac{n}{2}\rceil}^{\,n-2}
  \frac{\binom{\,s-1\,}{\,n-s-1\,}}{\binom{\,n-2\,}{s}}.
\end{align*}
    \end{lemma}
    
    \begin{IEEEproof}
    The possible start positions for any window are $\{0,1,\dots,n-\ell\}$. Any size-$s$ recovery set must include the extreme windows. Denote a recovery set \(\mathcal{R}\), and denote its corresponding set of start positions in an increasing order
    \[
    \mathcal{I_R}=\{x_1=0<x_2<\cdots<x_s=n-\ell\}.
    \]
    Let the gaps be $\delta_i=x_{i+1}-x_i\in\{1,\dots,\ell\}$ with $\sum_{i=1}^{s-1}\delta_i=n-\ell$, and set
    $\delta_i'=\delta_i-1\in\{0,\dots,\ell-1\}$. Then
    \[
    \sum_{i=1}^{s-1}\delta_i'=
    \underbrace{n-\ell-s+1}_{\text{telescopic}}.
    \]
    For each \(\mathcal{R}\in \mathrm{Rec}(\mathcal{A})\), \(\mathcal{I_R}\) corresponds to a unique set of gaps bijectively. Hence, $\alpha(s)$ equals the coefficient of $x^{\,n-\ell-s+1}$ in the generating function $\big(1+x+\cdots+x^{\ell-1}\big)^{s-1}$. Therefore, by a standard derivation, it can be shown that 
    \(
    \alpha(s)\tight\tight\tight
    \)\( \text{ }= \sum_{i=0}^{\min\!\left(s-1,\;\left\lfloor\frac{n-\ell-s+1}{\ell}\right\rfloor\right)}
    (-1)^i\binom{s-1}{i}\binom{\,n-\ell-i\ell-1\,}{\,s-2\,}\;
    .\)
    Note that \(\alpha(s)=0\) unless \begin{small}\(\lceil\tfrac{n}{\ell}\rceil \le s
    \)\end{small}, and note that for $\ell=2$, we get a simpler expression since $\delta_i'\in\{0,1\}$, and thus
    \(
    \;\alpha(s)=\binom{s-1}{\,n-s-1\,}\;
    \).
    \end{IEEEproof}
    
    Lemma~\ref{lem1} fully solves Problem~\ref{non cyclic windows problem} and gives an exact expression for the coverage time. Further, for the general case (arbitrary $\ell$), algebraic manipulation yields a more informative closed form.
    \begin{theorem}\label{non cyclic windows alternative form}
    For all integers $n>\ell\ge1$, it holds that
    \begin{align*}
    \mathbb{E}[T_{n,\ell{}}] 
    =& \tfrac{3}{2}(n - \ell + 1)
    \\&+\tight\tight\tight\tight\tight \sum_{i = 1}^{\lceil n / \ell \rceil - 2}\tight\tight\tight\tight \left( \tight\sum_{s = i + 1}^{n - \ell - i\ell + 1}\tight\tight\tight\tight\tight\tight
    (-1)^{i + 1} \small\binom{s - 1}{i}\tight \cdot\tight \frac{\binom{n - \ell - i\ell - 1}{s - 2}}{\binom{n - \ell}{s}}\tight\tight\tight \right)\tight.
    \end{align*}
    \end{theorem}

    \begin{IEEEproof}
    Denote \(c=\lceil \tfrac{n}{\ell} \rceil\). It holds that \(\tfrac{n}{c} \tight\leq \tight\ell\tight < \tight\tight\tfrac{n}{c - 1}\tight,\) and \(c \leq s \leq n - \ell\). Thus,
    \vspace{-1ex}
    \begin{align*}
    \begin{small}
    \left\lfloor\tight \tfrac{n - \ell - s + 1}{\ell}\tight \right\rfloor
    \tight\leq\tight \tfrac{n - \ell - s + 1}{\ell}\tight
    \leq\tight \tfrac{n - \frac{n}{c} - c + 1}{\frac{n}{c}}
    \tight=\tight c \tight-\tight 1\tight -\tight \tfrac{c(c + 1)}{n}
    \tight<\tight c\tight -\tight 1.
    \end{small}
    \end{align*}
    \vspace{-2ex}
    Therefore,
    \begin{small}
    \begin{align*}  
    \mathbb{E}[T_{n,\ell}] & \hspace{-0.25ex}=\hspace{-0.25ex} (n \hspace{-0.25ex}-\hspace{-0.25ex} \ell \hspace{-0.25ex}+\hspace{-0.25ex} 1) H_{n \tight -\tight  \ell \tight + \tight 1}
    \hspace{-0.25ex}-\hspace{-0.25ex} \sum_{s = c}^{n - \ell}\hspace{-0.25ex} \sum_{i = 0}^{c - 2}
    (-1)^i \hspace{-0.25ex}\binom{s - 1}{i}\hspace{-0.25ex} \frac{\binom{n - \ell - i\ell - 1}{s - 2}}{\binom{n - \ell}{s}} \\
    & = (n \tight- \tight\ell \tight+\tight 1)\tight H_{n \tight- \tight\ell \tight+ \tight1}\tight
    -\tight \sum_{i = 0}^{c - 2}\tight \sum_{s = c}^{n - \ell}
    \tight(-1)^i\tight \binom{s - 1}{i} \tight\frac{\binom{n - \ell - i\ell - 1}{s - 2}}{\binom{n - \ell}{s}}\hspace{-0.25ex},\vspace{-4ex}
    \end{align*}
    \end{small} 
    where the last equality holds by changing the order of summation.
    
    For all \( i \in \{0,1,\dots,c-2\} \), we must ensure that the binomial coefficient $\binom{n - \ell - i\ell - 1}{s - 2}$ does not vanish. This requires that
    \(
    s - 2 \leq n - \ell - i\ell - 1 \text{, or equivalently, } s \leq n - \ell - i\ell + 1.
    \) Therefore, we can write 
    \begin{align*}
    \mathbb{E}[T_{n,\ell}] =& (n - \ell + 1) \cdot H_{n - \ell + 1}
    \\&- \sum_{i = 0}^{c - 2} \sum_{s = c}^{\min\{n-\ell,n - \ell - i\ell+1\}}
    \tight\tight\tight\tight\tight\tight\tight\tight\tight\tight\tight\tight\tight\tight\tight\tight\tight\tight\tight\tight\tight(-1)^i \binom{s - 1}{i} \cdot \frac{\binom{n - \ell - i\ell - 1}{s - 2}}{\binom{n - \ell}{s}}.
    \end{align*}
Consider the sum,
    \begin{align*}
&\hspace{-4mm}\sum_{i=0}^{c-2}
  \sum_{s=c}^{\min\{n-\ell,n - \ell - i\ell+1\}}
  \tight\tight \tight \tight \tight \tight \tight \tight  \tight \tight \tight \tight \tight \tight \tight (-1)^i
  \binom{s-1}{i}
  \frac{\binom{n-\ell-i\ell-1}{s-2}}{\binom{n-\ell}{s}}\\
=& \tight\sum_{s=c}^{n-\ell}
  \tight\tight\frac{\binom{n-\ell-1}{s-2}}{\binom{n-\ell}{s}}
  \tight\tight+ \tight\tight\sum_{i=1}^{c-2}
  \tight\sum_{s=c}^{n\tight-\tight\ell\tight-\tight i\ell\tight+\tight 1\tight}
  \tight\tight\tight\tight\tight\tight(-1)^i
  \binom{s-1}{i}
  \frac{\binom{n-\ell-i\ell-1}{s-2}}{\binom{n-\ell}{s}}\\
=& \sum_{s=2}^{n-\ell}
  \frac{\binom{n-\ell-1}{s-2}}{\binom{n-\ell}{s}}
  - \sum_{s=2}^{c-1}
  \frac{\binom{n-\ell-1}{s-2}}{\binom{n-\ell}{s}}
  \\& + \sum_{i=1}^{c-2}
  \sum_{s=c}^{n-\ell-i\ell+1}
  (-1)^i
  \binom{s-1}{i}
  \frac{\binom{n-\ell-i\ell-1}{s-2}}{\binom{n-\ell}{s}}.
\end{align*}
It can be shown algebraically that  
\begin{align*}
    \sum_{s = 2}^{n - \ell} \frac{\binom{n - \ell - 1}{s - 2}}{\binom{n - \ell}{s}} = (n - \ell + 1)\left( H_{n - \ell + 1} - \tfrac{3}{2} \right).
    \end{align*}
    Therefore, the sum becomes
    \begin{align*}
    &(n - \ell + 1)\left( H_{n - \ell + 1} - \tfrac{3}{2} \right)
    - \sum_{s = 2}^{c - 1} \frac{\binom{n - \ell - 1}{s - 2}}{\binom{n - \ell}{s}}
    \\& + \sum_{i = 1}^{c - 2} \sum_{s = c}^{n - \ell - i\ell + 1} (-1)^i \binom{s - 1}{i} \cdot \frac{\binom{n - \ell - i\ell - 1}{s - 2}}{\binom{n - \ell}{s}}.\end{align*}
    Substituting back into the expectation, we get,
    \begin{align*}
    \mathbb{E}[T_{n,\ell}]=& (n - \ell + 1) \cdot H_{n - \ell + 1}
    \\& - \Bigg[ (n - \ell + 1)\left( H_{n - \ell + 1} - \tfrac{3}{2} \right)
    - \sum_{s = 2}^{c - 1} \frac{\binom{n - \ell - 1}{s - 2}}{\binom{n - \ell}{s}} 
    \\&+ \sum_{i = 1}^{c - 2} \sum_{s = c}^{n - \ell - i\ell + 1} (-1)^i \binom{s - 1}{i} \cdot \frac{\binom{n - \ell - i\ell - 1}{s - 2}}{\binom{n - \ell}{s}} \Bigg] 
    \\=& \tfrac{3}{2}(n - \ell + 1)
    + \sum_{s = 2}^{c - 1} \frac{\binom{n - \ell - 1}{s - 2}}{\binom{n - \ell}{s}}
    \\&- \sum_{i = 1}^{c - 2} \sum_{s = c}^{n - \ell - i\ell + 1} (-1)^i \binom{s - 1}{i} \cdot \frac{\binom{n - \ell - i\ell - 1}{s - 2}}{\binom{n - \ell}{s}}.
\end{align*}
Further, by adding and subtracting the term 
\vspace{-1ex}
\begin{align*}
    \sum_{i = 1}^{c - 2} \sum_{s = i + 1}^{c - 1} (-1)^i \binom{s - 1}{i} \cdot \frac{\binom{n - \ell - i\ell - 1}{s - 2}}{\binom{n - \ell}{s}}, 
    \end{align*} 
    \vspace{-0.5ex}we get
    \vspace{-1ex}
\begin{align*}
    \E[T_{n,\ell}]
    = & \tfrac{3}{2}(n - \ell + 1)
      + \sum_{s = 2}^{c - 1} \frac{\binom{n - \ell - 1}{s - 2}}{\binom{n - \ell}{s}}
     \\ & +\sum_{i = 1}^{c - 2} \sum_{s = i + 1}^{c - 1} (-1)^i \binom{s - 1}{i} \cdot \frac{\binom{n - \ell - i\ell - 1}{s - 2}}{\binom{n - \ell}{s}} 
    \\& - \tight \sum_{i = 1}^{c - 2} \tight\tight \sum_{s = i + 1}^{n \tight-\tight \ell\tight -\tight i\ell  \tight +\tight 1}\tight\tight (-1)^i \tight\binom{\tight s - 1\tight}{i}\tight \cdot\tight \frac{\binom{\tight n \tight-\tight \ell \tight -\tight i\ell \tight - 1 \tight}{\tight s - 2 \tight}}{\binom{n - \ell}{s}}
    \tight \\
    = &\tfrac{3}{2}(\tight n\tight -\tight \ell \tight+\tight 1)
      \\&+\tight \tight \left[
          \tight\sum_{s = 2}^{c - 1}\tight\tight \frac{\binom{n - \ell - 1}{s - 2}}{\binom{n - \ell}{s}}
          \tight\tight+\tight\tight \sum_{i = 1}^{c - 2}\tight\tight\tight\tight \sum_{s = i + 1}^{c - 1} \tight\tight(-1)^i \binom{s - 1}{i} \tight\tight\cdot\tight\tight \frac{\binom{n - \ell - i\ell - 1}{s - 2}}{\binom{n - \ell}{s}}
        \tight\right]
        \\&- \sum_{i = 1}^{c - 2} \sum_{s = i + 1}^{n - \ell - i\ell + 1} (-1)^i \binom{s - 1}{i} \cdot \frac{\binom{n - \ell - i\ell - 1}{s - 2}}{\binom{n - \ell}{s}}.\end{align*}
    Observe that the sum inside the square brackets is equal to zero. Indeed,
    \vspace{-1ex}
    \begin{small}
    \begin{flalign*}
&\hspace{1em}\sum_{s = 2}^{c - 1}
  \frac{\binom{n - \ell - 1}{s - 2}}{\binom{n - \ell}{s}}
  + \sum_{i = 1}^{c - 2}
    \sum_{s = i + 1}^{c - 1}
    (-1)^i \binom{s - 1}{i}
    \frac{\binom{n - \ell - i\ell - 1}{s - 2}}{\binom{n - \ell}{s}}\\&= \sum_{s\tight=\tight2}^{c\tight-\tight1}
    (-1)^{0}\tight\binom{s\tight-\tight1}{0}
    \tight\frac{\binom{n\tight-\tight\ell\tight-\tight1}{s\tight-\tight2}}{\binom{n\tight-\tight\ell}{s}}
  \tight+\tight\sum_{i\tight=\tight1}^{c\tight-\tight2}
    \sum_{s\tight=\tight i\tight+\tight1}^{c\tight-\tight1}
    (-1)^{i}\tight\binom{s\tight-\tight1}{i}
    \tight\frac{\binom{n\tight-\tight\ell\tight-\tight i\ell\tight-\tight1}{s\tight-\tight2}}{\binom{n\tight-\tight\ell}{s}}\\
&=\sum_{s\tight=\tight1}^{c\tight-\tight1}
    (-1)^{0}\tight\binom{s\tight-\tight1}{0}
    \tight\frac{\binom{n\tight-\tight\ell\tight-\tight1}{s\tight-\tight2}}{\binom{n\tight-\tight\ell}{s}}
  \tight+\tight\sum_{i\tight=\tight1}^{c\tight-\tight2}
    \sum_{s\tight=\tight i\tight+\tight1}^{c\tight-\tight1}
    (-1)^{i}\tight\binom{s\tight-\tight1}{i}
    \tight\frac{\binom{n\tight-\tight\ell\tight-\tight i\ell\tight-\tight1}{s\tight-\tight2}}{\binom{n\tight-\tight\ell}{s}}
\\
&= \sum_{i = 0}^{c - 2}
    \sum_{s = i + 1}^{c - 1}
    (-1)^i \binom{s - 1}{i}
    \frac{\binom{n - \ell - i\ell - 1}{s - 2}}{\binom{n - \ell}{s}}\\
&= \sum_{s = 1}^{c - 1}
    \sum_{i = 0}^{s - 1}
    (-1)^i \binom{s - 1}{i}
    \frac{\binom{n - \ell - i\ell - 1}{s - 2}}{\binom{n - \ell}{s}}\\
&= \sum_{s = 1}^{c - 1}
    \frac{1}{\binom{n - \ell}{s}}
    \sum_{i = 0}^{s - 1}
    (-1)^i \binom{s - 1}{i}
    \binom{n - \ell - i\ell - 1}{s - 2}. &&
\end{flalign*}
\end{small}
    \hspace{-5pt}For any fixed \( s \in \{1,2,\dots,c-1\} \), the inner sum is an inclusion-exclusion formula for counting the number of binary vectors of length \( n - \ell - 1 \) and weight exactly \( s - 2 \), such that for each \( i \in \{1,\dots,s - 1\} \), there is at least one bit of value one among the bits in positions 
\(\{(i - 1)\ell + 1, \dots, i\ell\}
    \). This would require the vector to have at least \( s - 1\) nonzero entries (i.e., weight at least \( s \)), which contradicts the condition that the weight is \( s - 2 \). Therefore, the number of options is zero. Hence the entire bracketed sum vanishes. This concludes that the expectation is
    \begin{align*}
    \mathbb{E}[T_{n,\ell}] =& \tfrac{3}{2}(n - \ell + 1)
    \\&+ \sum_{i = 1}^{c - 2} \left( \sum_{s = i + 1}^{n - \ell - i\ell + 1}
    \tight\tight\tight\tight(-1)^{i + 1} \binom{s - 1}{i} \cdot \frac{\binom{n - \ell - i\ell - 1}{s - 2}}{\binom{n - \ell}{s}} \right).
    \end{align*}
    Recall that \(c=\lceil\frac{n}{\ell}\rceil\), can be substituted back and yield the theorem's statement.
    \end{IEEEproof}
    Thus, a final expression for the expectation is obtained. 
    In Section~\ref{cyclic vs continous}, we show that for $\ell=\omega(\log \left( n \right))$, the double sum \(S_{n,\ell}\) satisfies
    \(S_{n,\ell} =o\big(\tfrac{3}{2}(n-\ell+1)\big)\), 
    establishing that the expectation is dominated by its linear term (extremes coverage), and for $\ell=o\left(\log(n) \right),$ $\mathbb{E}[T_{n,\ell{}}]\approx \frac{n}{\ell}\log \left( \frac{n}{\ell} \right)$ (middle coverage), leaving only the $\ell=\Theta\left(\log(n)\right)$ regime's asymptotics ambiguous. 

    \subsection{Large Windows}
This section presents the expected coverage time for special regimes of $\ell$ relative to $n$.     
    \begin{theorem}\label{thm:large-windows}
    For all integers $\frac{n}{2}\le \ell< n$, it holds that
    \[
    \;\mathbb{E}[T_{n,\ell}] \;=\; \tfrac{3}{2}\,\big(n-\ell+1\big). \;
    \]
    \end{theorem}
    
    \begin{IEEEproof}
    Write $M=n-\ell+1$ for the number of windows. When $\ell\ge n/2$, the two extreme windows,
    $\mathrm{win}_\ell(0)=[0,\ell-1]$ and $\mathrm{win}_\ell(n-\ell)=[n-\ell,n-1]$,
    already cover all positions (each endpoint $0$ and $n{-}1$ belongs to exactly one of them).
    Hence full coverage occurs if and only if both extreme windows have appeared. Draws are i.i.d.\ uniform over the $M$ windows. The waiting time to see either extreme window is
    geometric with success probability $2/M$, of expectation $M/2$. After the first extreme window is seen,
    the waiting time to see the other extreme window is geometric with success probability $1/M$, of
    expectation $M$. By linearity of expectation,
    \[
    \mathbb{E}[T_{n,\ell}] \;=\; \frac{M}{2} + M \;=\; \tfrac{3}{2}\,M
    \;=\; \tfrac{3}{2}\,(n-\ell+1).\qedhere
    \]
    \end{IEEEproof}

    \begin{theorem}\label{thm:third-to-half}
    For all integers $\frac{n}{3}\le\ell <\tfrac{n}{2}  $, it holds that
    \[
    \;
    \mathbb{E}[T_{n,\ell}]
    = \tfrac{3}{2}\,\big(n-\ell+1\big)
    \;+\;
    \big(n-\ell+1\big)\,\tfrac{6n-10\ell}{\ell(\ell+1)(\ell+2)(\ell+3)}
    \;.
    \]
    \end{theorem}
    \begin{IEEEproof}
We first observe that full coverage occurs if and only if both extreme windows were sampled and all $n-2\ell$ middle positions are covered. So only coverage of the middle \(n-2\ell+2\) positions matters.
We model the process using a three-layered Markov chain. States are denoted by \( S_i^k \), where \( i \in \{0, 1, \ldots, n-2\ell\} \) counts the number of unseen middle positions, and \( k \in \{0, 1, 2\} \) indicates how many extreme windows have been observed. Note that the uncovered middle positions always appear continuously. For each state $S_i^k$, the number of windows that cover all $i$ uncovered middle positions is $\ell-i+1$, giving the transition probability $\Pr[S_i^k\rightarrow S_0^k]=\tfrac{\ell-i+1}{n-\ell+1}$. For each $1\le j<i$, the number of windows covering exactly $i-j$ new middle positions is exactly 2, giving a transition probability $\Pr[S_i^k\rightarrow S_j^k]=\tfrac{2}{n-\ell+1}$. The number of windows that reveal an uncovered extreme window is $2-k$, giving a transition probability $\Pr[S_i^k\rightarrow S_i^{k+1}]=\tfrac{2-k}{n-\ell+1}$. Since all probabilities sum to $1$, the transition probability of no new covered positions is $\Pr[S_i^k\rightarrow S_i^k]=\tfrac{n-2\ell-i+k}{n-\ell+1}$. We solve the markov chain in a bottom up, as can be demonstrated visually in Figure~\ref{fig:3_layer_markov}.

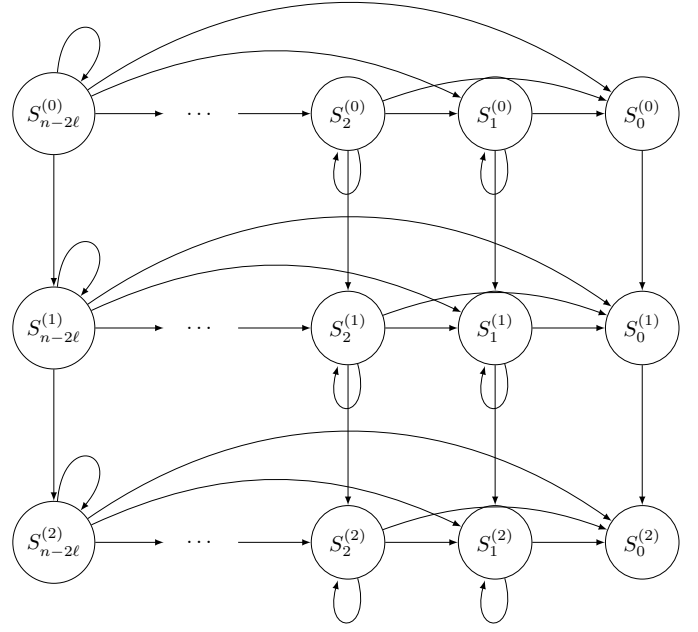
\begin{figure}[t]
    \centering
  \resizebox{0.5\textwidth}{!}{
\begin{tikzpicture}[->, >=latex, node distance=2.2cm and 2.4cm, on grid, auto,
every node/.style={circle, draw, minimum size=1.2cm, align=center}]

\node (n2m0) {$S_{n-2\ell}^{(0)}$};
\node (dots0) [right=of n2m0, draw=none] {$\dots$};
\node (s2_0) [right=of dots0] {$S_2^{(0)}$};
\node (s1_0) [right=of s2_0] {$S_1^{(0)}$};
\node (s0_0) [right=of s1_0] {$S_0^{(0)}$};

\node (n2m1) [below=3.5cm of n2m0] {$S_{n-2\ell}^{(1)}$};
\node (dots1) [right=of n2m1, draw=none] {$\dots$};
\node (s2_1) [right=of dots1] {$S_2^{(1)}$};
\node (s1_1) [right=of s2_1] {$S_1^{(1)}$};
\node (s0_1) [right=of s1_1] {$S_0^{(1)}$};

\node (n2m2) [below=3.5cm of n2m1] {$S_{n-2\ell}^{(2)}$};
\node (dots2) [right=of n2m2, draw=none] {$\dots$};
\node (s2_2) [right=of dots2] {$S_2^{(2)}$};
\node (s1_2) [right=of s2_2] {$S_1^{(2)}$};
\node (s0_2) [right=of s1_2] {$S_0^{(2)}$};

\draw[->] (n2m0) -- (dots0);
\draw[->] (dots0) -- (s2_0);
\draw[->] (s2_0) -- (s1_0);
\draw[->] (s1_0) -- (s0_0);

\draw[->, bend left=25] (n2m0) to (s1_0);
\draw[->, bend left=35] (n2m0) to (s0_0);
\draw[->, bend left=20] (s2_0) to (s0_0);

\draw[->] (n2m1) -- (dots1);
\draw[->] (dots1) -- (s2_1);
\draw[->] (s2_1) -- (s1_1);
\draw[->] (s1_1) -- (s0_1);

\draw[->, bend left=25] (n2m1) to (s1_1);
\draw[->, bend left=35] (n2m1) to (s0_1);
\draw[->, bend left=20] (s2_1) to (s0_1);

\draw[->] (n2m2) -- (dots2);
\draw[->] (dots2) -- (s2_2);
\draw[->] (s2_2) -- (s1_2);
\draw[->] (s1_2) -- (s0_2);

\draw[->, bend left=25] (n2m2) to (s1_2);
\draw[->, bend left=35] (n2m2) to (s0_2);
\draw[->, bend left=20] (s2_2) to (s0_2);

\draw[->] (n2m0) -- (n2m1);
\draw[->] (s2_0) -- (s2_1);
\draw[->] (s1_0) -- (s1_1);
\draw[->] (s0_0) -- (s0_1);

\draw[->] (n2m1) -- (n2m2);
\draw[->] (s2_1) -- (s2_2);
\draw[->] (s1_1) -- (s1_2);
\draw[->] (s0_1) -- (s0_2);

\draw[->] (n2m0) edge[loop , min distance=12mm, in=50, out=85] (n2m0);
\draw[->] (s2_0) edge[loop below] (s2_0);
\draw[->] (s1_0) edge[loop below] (s1_0);

\draw[->] (n2m1) edge[loop , min distance=12mm, in=50, out=85] (n2m1);
\draw[->] (s2_1) edge[loop below] (s2_1);
\draw[->] (s1_1) edge[loop below] (s1_1);

\draw[->] (n2m2) edge[loop , min distance=12mm, in=50, out=85] (n2m2);
\draw[->] (s2_2) edge[loop below] (s2_2);
\draw[->] (s1_2) edge[loop below] (s1_2);

\end{tikzpicture}
}
\caption{Layered Markov chain, representing the case \(\ell\ge \frac{n}{3}\).}
\label{fig:3_layer_markov}
\end{figure}    
    For $k = 2$, using the transition probabilities, we derive the following recursive relation,
$$\E[S_i^2 \tight\tight\tight\to\tight\tight\tight S_0^2]\tight \tight=\tight\tight 1\tight \tight+\tight\tight \tfrac{n - 2\ell - i + 2}{n - \ell + 1} \E[S_i^2 \tight\tight\tight\to\tight\tight\tight S_0^2]\tight \tight+\tight\tight \tfrac{2}{n - \ell + 1} \tight\tight\tight\sum_{j = 1}^{i - 1} \tight\E[S_j^2 \tight\tight\tight\to\tight\tight\tight S_0^2],$$
and hence 
$$\E[S_i^2 \to S_0^2] = \tfrac{n - \ell + 1}{\ell - 1 + i} + \tfrac{2}{\ell - 1 + i} \sum_{j = 1}^{i - 1} \E[S_j^2 \to S_0^2].$$
And it can be shown by induction that
    \[
    E[S_i^2 \to S_0^2] = \tfrac{n - \ell + 1}{\ell} + \frac{n - \ell + 1}{\ell(\ell + 1)}(i - 1).
    \]
    Similarly, for \( k = 1 \), it can be derived that
\[
    \E[S_i^1\tight\tight\tight \to\tight\tight\tight S_0^2] \tight\tight=\tight \tight\tfrac{(\ell - i + 2)(n - \ell + 1)}{\ell + i} \tight\tight+\tight\tight \tfrac{n - \ell + 1}{\ell(\ell + 1)} \tight\tight+\tight\tight \tfrac{2}{\ell + i}\tight\tight\tight \sum_{j = 1}^{i - 1} \tight \E[S_j^1\tight\tight\tight \to\tight\tight\tight S_0^2],
    \]
    and by induction we get
    \[
    \E[S_i^1 \tight\tight\tight\to\tight\tight\tight S_0^2] = (n\tight -\tight \ell \tight+\tight 1)\tight +\tight \tfrac{n - \ell + 1}{\ell(\ell + 1)}\tight +\tight \tfrac{2(n - \ell + 1)}{\ell(\ell + 1)(\ell + 2)} (i - 1).
    \]
    For $k=0$, the relation becomes
    \begin{align*}
    \E[S_i^0 \tight\tight\tight\to\tight\tight\tight S_0^2]\tight =&\tight \tfrac{n - \ell + 1}{\ell + 1 + i}\tight +\tight \tfrac{\ell - i + 1}{\ell + 1 + i} \tight\cdot\tight \tfrac{3(n - \ell + 1)}{2}
    \tight\\& +\tight\tight\tight \tfrac{2}{\ell\tight +\tight 1\tight +\tight i}\tight\tight \left( \tight\tight\tight\small(n\tight \tight-\tight\tight \ell\tight\tight +\tight\tight 1\tight) \tight\tight  +\tight\tight \tfrac{n - \ell + 1}{\ell(\ell + 1)}\tight \tight+\tight\tight \tfrac{2(n - \ell + 1)}{\ell(\ell + 1)(\ell + 2)}(i \tight-\tight 1) \tight\tight\tight\right)
    \tight\\&+\tight \tfrac{2}{\ell + 1 + i} \sum_{j = 1}^{i - 1} \E[S_j^0 \to S_0^2].
    \end{align*}
    And a final induction argument gives
    \begin{align*}
    \E[S_i^0 \tight\tight\tight\to\tight\tight\tight S_0^2] \tight=\tight \tfrac{3(n - \ell + 1)}{2}\tight +\tight \tfrac{2(n - \ell + 1)}{\ell(\ell + 1)(\ell + 2)}\tight +\tight \tfrac{6(n - \ell + 1)}{\ell(\ell + 1)(\ell + 2)(\ell + 3)} (i - 1).
    \end{align*}
    Note that \(\mathbb{E}[T_{n,\ell}] = \E[S_{n - 2\ell}^0 \tight \to\tight S_0^2]\), therefore, substituting into the closed-form expression yields
    \[\tight
    \mathbb{E}[T_{n,\ell}] \tight=\tight \tfrac{3(n - \ell + 1)}{2} \tight+\tight \tfrac{2(n - \ell + 1)}{\ell(\ell + 1)(\ell + 2)} \tight+\tight \tfrac{6(n - \ell + 1)(n - 2\ell - 1)}{\ell(\ell + 1)(\ell + 2)(\ell + 3)}.
    \]
    Rearranging terms yields
    the expression in the theorem.
    \end{IEEEproof}
    Note that this result also gives the expected number of steps required for getting from any $i\in[n]$ uncovered positions, to $j$.

    After analyzing two specific cases representing the scenarios of large windows ($\ell\ge\tfrac{n}{3}$), lastly an analysis of a more general result is given in the next theorem.   
\begin{theorem}
\label{thm:large-ell-asymptotic}
    For any $\ell \geq \frac{n}{c}$ with a constant $c > 1$, it holds that
    \[
    \E[T_{n,\ell}] = \tfrac{3}{2}(n - \ell + 1)+ O\Big(\tfrac{1}{n^2}\Big).
    \]
    \end{theorem}
    
    \begin{IEEEproof}
    Denote \(c=\lceil \frac{n}{\ell}\rceil\). By Theorem~\ref{non cyclic windows alternative form},
    \[
    \mathbb{E}[T_{n,\ell}]\tight =\tight \tfrac{3}{2}(n - \ell + 1)
    \tight+\tight \tight\sum_{i = 1}^{c - 2}\tight \sum_{s = i + 1}^{n - \ell - i\ell + 1}
    \tight\tight\tight\tight\tight(-1)^{i + 1} \tight\small\binom{s - 1}{i}\tight \cdot\tight
    \tfrac{\binom{n - \ell - i\ell - 1}{s - 2}}{\binom{n - \ell}{s}}.
    \]
    The proof proceeds by bounding the absolute sum.
    It holds that
    \begin{align*}
    \sum_{i = 1}^{c - 2}\tight &\sum_{s = i + 1}^{n - \ell - i\ell + 1}
    \tight\tight\tight\tight\tight(-1)^{i + 1} \tight\small\binom{s - 1}{i}\tight \cdot\tight
    \tfrac{\binom{n - \ell - i\ell - 1}{s - 2}}{\binom{n - \ell}{s}}
    \\&\overset{(a)}{\le}
 (c-2)\tight\tight\tight\tight\tight\tight\sum_{s = 2}^{n - 2\ell + 1} s^{c-2}
    \cdot \tfrac{\binom{n - 2\ell - 1}{s - 2}}{\binom{n - \ell}{s}} \\&= (c\tight-\tight2)\tight\tight\tight\tight\tight\tight\tight\sum_{s = 2}^{n - 2\ell + 1} \tight\tight\tight\tight\tight s^{c-2}\tight
    \tight\cdot\tight\tight \tfrac{s(s - 1)}{(n-\ell)(n-\ell-1)}
 \tight\tight   \prod_{j = 1}^{s - 2}
    \tight\tight\big(1\tight-\tight\tight\tfrac{\ell-1}{n - \ell - j - 1}\big)
    \\& \overset{(b)}{\le}
 \tfrac{(c-2)}{(n-\ell-1)^2}\sum_{s = 2}^{n - 2\ell + 1} s^{c}
    \cdot \prod_{j=1}^{s-2}
    e^{-\tfrac{\ell-1}{n-\ell-j-1}}
    \\&= \tfrac{(c-2)}{(n-\ell-1)^2}
    \sum_{s = 2}^{n - 2\ell + 1} s^{c}\cdot e^{\left(
    -\sum_{j=1}^{s-2}\frac{\ell-1}{\,n-\ell-j-1\,}
    \right)} 
    \\&\overset{(c)}{\le}
 \tfrac{(c-2)}{(n-\ell-1)^2} \sum_{s=2}^{n-2\ell+1} s^{c}\,
       e^{-\,\frac{(\ell-1)(s-2)}{\,n-\ell\,}}  
    \\&= \tfrac{(c-2)e^{2\alpha}}{(n-\ell-1)^2}
    \sum_{s=2}^{n-2\ell+1} s^{c}\, e^{-\frac{\ell-1}{n-\ell} s}
    \\&
\overset{(d)}{\le}
 \tfrac{(c-2)e^{2}}{(n-\ell-1)^2}
    \sum_{s=2}^{\infty} s^{c}\, e^{-\tfrac{1}{2(c-1)} s}
    = O\!\big(\tfrac{1}{n^2}\big).
    \end{align*} Where $(a)$ is true since $i\le c-2$, $(b)$ is true since $1-x\le e^{-x}$, $(c)$ is true since we took the maximal term and $(d)$ is true since $\tfrac{\ell -1}{\,n-\ell\,}\ge
  \tfrac{1}{2(c-1)}$, and since $\sum_{s=1}^\infty{s^a}\cdot e^{-bs}$ converges for any constants $a,b>0$. 
    \end{IEEEproof}

    \section{Problem \ref{cyclic windows problem}: The Cyclic Windows Model}
    \label{sec:cyclic wondows - solution}
    
    This section analyses Problem \ref{cyclic windows problem}, the cyclic version of the windows coverage problem. First, a general solution is being presented. By applying Section~\ref{anina generalization}'s approach, we get that in the cyclic case, the hypergraph is
    \[
    \mathcal H^{\circlearrowright}_{n,\ell}=(V,E),\qquad
    V=[n],\qquad
    E=\mathcal{A}=\ \mathcal{W}_{n,\ell}^{\circlearrowright}.
    \]
    $N=|E|=n$. Every position lies in exactly $\ell$ cyclic windows, hence the largest non–recovery set has size \(M=n-\ell \quad\text{(omit all $\ell$ windows covering one fixed position).}\)
    Let $\alpha^{\circlearrowright}(s)$ be the number of cyclic recovery sets of size $s$. As in Problem~\ref{non cyclic windows problem}, 
    $\alpha^{\circlearrowright}(s)=0$ unless $s\ge\lceil \frac{n}{\ell}\rceil$. Therefore, by Theorem~\ref{thm:hypergraph-expectation},
    \begin{equation}\label{eq:ET-cyclic}
    \mathbb{E}\!\big[T_{n,\ell}^\circlearrowright\big]
    = n\big(H_n-H_{\ell-1}\big)
    -\sum_{s=\lceil \frac{n}{\ell}\rceil}^{\,n-\ell}
    \frac{\alpha^{\circlearrowright}(s)}{\binom{n-1}{s}}.
    \end{equation}
    In the same manner, the task reduces to determining $\alpha^{\circlearrowright}(s)$, yielding the following.
    \begin{theorem}\label{thm:general-cyclic}
    For all integers $n>\ell\ge1$, it holds that
    \begin{align*}
    \mathbb{E}\!\big[T_{n,\ell}^\circlearrowright\big]
    =& \hspace{1mm}  n\big(H_n-H_{\ell-1}\big)
    \\&-n\sum_{s=\lceil \frac{n}{\ell} \rceil}^{\,n-\ell}\frac{1}{s}\sum_{j=0}^{\left\lfloor\frac{n-s}{\ell}\right\rfloor}
    (-1)^j
     \binom{s}{j}\cdot \frac{\,\binom{\,n-\ell j-1\,}{\,s-1\,}}{\binom{n-1}{s}}.
     \\\text{In particular,}        \\&\hspace{-4em}\mathbb{E}\!\big[T_{n,2}^\circlearrowright\big]
    =
    n(H_n-1)\;-\!
    \sum_{s=\lceil \frac{n}{2} \rceil}^{\,n-2}
    \frac{\binom{s-1}{\,n-s-1\,}+\binom{s}{\,n-s\,}}{\binom{n-1}{s}}.
    \end{align*}
    \end{theorem}
    \begin{IEEEproof}
    We calculate the value of $\alpha^{\circlearrowright}(s)$ for $s\geq \left\lceil\frac{n}{\ell}\right\rceil$ and we first consider the case $\ell=2$. Denote a recovery set $\mathcal{R}$. Let $w^\ast=\{n{-}1,0\}$ be the wrap–around window. We split according to whether $w^\ast\in\mathcal R$. In case $w^\ast\notin\mathcal R$, the problem is exactly as in Problem \ref{non cyclic windows problem}, hence we have \(\binom{s-1}{\,n-s-1\,}\) recovery sets not containing $w^*$. In the case where $w^\ast\in\mathcal R$, $\{1,2,\dots,n-2\}$ must be covered using $s-1$ windows from the remaining $n-1$ edges 
    $\{0,1\},\{1,2\},\dots,\{n-2,n-1\}$. Denote $\mathcal{R}$'s corresponding set of start positions by
    \[
    \mathcal{I_R}=\{x_1<x_2<\cdots<x_{s-1}\}.
    \]
    For full coverage, the extremes must satisfy both \(x_1\in\{0,1\}\) and \(x_{s-1}\in\{n-3,n-2\} \), so there are exactly $4$ endpoint options. As in Problem~\ref{non cyclic windows problem}, set the gaps
    \(
    \delta_i=x_{i+1}-x_i\in\{1,2\}\quad(1\le i\le s-2),\) and 
    
    $$\sum_{i=1}^{s-2}\delta_i
    =\underbrace{x_{s-1}-x_1}_{\text{telescopic}}.$$
    Let $\delta'_i=\delta_i-1\in\{0,1\}$. Then the number of recovery sets is the number of solutions to \(
    \sum_{i=1}^{s-2}\delta'_i=x_{s-1}-x_1-s+2.\)
    For the four endpoint choices $(x_1,x_{s-1})\in\{0,1\}\times\{n-3,n-2\}$ the right–hand side equals,
    respectively,
    \[
    n-s,\quad n-s-1,\quad n-s-1,\quad n-s-2.
    \]
    Since $\delta'\in\{0,1\}$ the four possibilities give
    \(
    \binom{s-2}{\,n-s\,}+2\binom{s-2}{\,n-s-1\,}+\binom{s-2}{\,n-s-2\,}
    =\binom{s}{\,n-s\,},
    \) recovery sets altogether, where the last identity can be proved combinatorially as follows. Consider the number of length-$s$ binary vectors of Hamming weight $n-s$. The right hand side is clearly the solution to this problem. As for the left side, we count the vectors by 4 mutually disjoint groups with the bits in positions 1 and 2 being $(0,0),(0,1),(1,0),(1,1)$. Combining the two cases, the number of recovery sets is
    \(\alpha^\circlearrowright(s)=
    \binom{s-1}{\,n-s-1\,}+\binom{s}{\,n-s}.
    \)
    Substituting into Equation~(\ref{eq:ET-cyclic}) gives the result. This concludes the solution for the case $\ell=2$. 
    
    We now move on to the general case. A size-$s$ recovery set is specified by its $s$ start indices on the circle. Since there is no canonical ``first'' index, we fix the rotation by anchoring one chosen start index at $0$. Denote such recovery set as $\mathcal{R}$,  and denote its corresponding set of start positions
    \(
    \mathcal{I_R}=\{x_1=0<x_2<\cdots<x_s<n\}\subseteq\{0,1,\dots,n-1\}.
    \) Focusing on only a subset of recovery sets, all those which contain $0$ as one of the starting positions. Define the \emph{cyclic gaps}
    \(
    \delta_1=x_2-x_1,\quad \delta_2=x_3-x_2,\ \dots,\ \delta_{s-1}=x_s-x_{s-1},\quad \delta_s=n-x_s.
    \)
    As in Problem~\ref{non cyclic windows problem}, the number of recovery sets is the number of solutions to \(
    \sum_{i=1}^{s}\delta_i
    =\underbrace{(x_2-x_1)+\cdots+(x_s-x_{s-1})+(n-x_s)}_{\text{telescopic}}
    = n\), with \(1\le \delta_i\le \ell\).
    Denote $\delta_i'=\delta_i-1\in\{0,1,\dots,\ell-1\}$. Then
    \(
    \sum_{i=1}^{s}\delta_i' = n-s
    \). Let $\beta^{\circlearrowright}(s)$ be the number of solutions, which is the number of cyclic recovery sets that contain $0$. Shortly, like in Problem~\ref{non cyclic windows problem}, the solution is derived from the generating function

    \vspace{-3ex}
    \begin{small}
\begin{align*}
\beta^{\circlearrowright}(s)
\tight\tight=\tight\tight\tight\big[x^{\,n\tight-\tight s}\big]\tight
\tight(1\tight+\tight x\tight+\tight\cdots\tight+\tight x^{\ell\tight-\tight1}\tight)^{s}
\tight\tight=\tight\tight\tight\sum_{j\tight=\tight0}^{\left\lfloor\tfrac{n\tight-\tight s}{\ell}\right\rfloor}
\tight\tight\tight\tight(-1)^{j}\tight\tight\binom{s}{j}\tight\tight\tight\binom{\,n\tight-\tight\ell j\tight-\tight1\,}{\,s\tight-\tight1\,}\tight\tight.
\end{align*}
    \end{small}
    \hspace{-5pt}To compute the number of recovery sets, a method similar to the one in the proof of Burnside's Lemma for finite groups (\tight\cite{burnside1897}) is being used. Consider the set of pairs
    \(
    \mathcal{P}_s=\{(\mathcal{R},i):\ \mathcal{R} \text{ is a size-$s$ cyclic recovery set},\ i\in \mathcal{I_R}\}.
    \)
    The size of $\mathcal{P}_s$ can be counted in two different ways. It can be counted looking from the perspective of the sets. Each $R$ contributes exactly $s$ pairs to $\mathcal{P}_s$, so $|\mathcal {P}_s|=s\cdot\,\alpha^{\circlearrowright}(s)$. It can be also counted, looking from the perspective of the indices. Since the setup is symmetrical with respect to cyclic rotation, for any $i\in[n]$ the number of pairs it appears in is the same. Counting the number of times 0 appears gives us exactly $\beta^{\circlearrowright}(s)$. Hence, $|\mathcal {P}_s|=n\cdot\,\beta^{\circlearrowright}(s)$. Therefore, we get
    \(
    \alpha^{\circlearrowright}(s)
    =\frac{n}{s}\,\beta^{\circlearrowright}(s)
    \)\(
    =\frac{n}{s}\sum_{j=0}^{\left\lfloor\frac{n-s}{\ell}\right\rfloor}
    (-1)^j\binom{s}{j}\binom{\,n-\ell j-1\,}{\,s-1\,}.\;
    \)
    Substituting into Equation~\ref{eq:ET-cyclic}, the final result is obtained.
    \end{IEEEproof}
    In Section~\ref{cyclic vs continous}, for $\ell=o(n)$, the asymptotic behavior of $
    \mathbb{E}\!\big[T_{n,\ell}^\circlearrowright\big]\approx\frac{n}{\ell} \log (\tfrac{n}{\ell})$ is discussed, showing its connection to Problem~\ref{continous problem}. In Section \ref{subsets and monotonicity section} a monotonicity property is presented, regarding a family of sub models that Problem \ref{cyclic windows problem} induces and their multi-dimensional extensions.

    \section{Deriving Results and Bounds from Problem \ref{continous problem}, The Continuous Circle}\label{cyclic vs continous}
    Problems \ref{cyclic windows problem} and \ref{continous problem} share a common structure. In both problems, arcs on a circle are being sampled until full coverage of the circumference. The two models can be compared and it is now shown that the discrete model has an expected number of samples that is at most the expected number in the continuous model.
    
    \begin{lemma}
    \label{lem:cyc-vs-cont}
    For all $1\leq \ell \leq n$, it holds that
    \[
    \mathbb{E}\!\big[T_{n,\ell}^\circlearrowright\big]\;\le\; \E[\mathcal{T}_\frac{\ell}{n}].
    \]
    \end{lemma}
    
    \begin{IEEEproof}
    The idea is that $T_{n,\ell}$'s distribution can be created via $\mathcal{T}_\frac{\ell}{n}$'s distribution, and it can be shown that for every sampled set of arcs that cover the circle, the corresponding created windows, cover the length-$n$ circle, thus showing a stochastic dominance of $\mathcal{T}_{\frac{\ell}{n}}$ over $T_{n,\ell}$. 
    
    Let $C_i \sim \mathrm{Unif}[0,1)$, and define \(D_i = \lfloor n C_i \rfloor \in \{0,1,\dots,n-1\}.\) We interpret $D_i$ as the starting index of a cyclic window of length $\ell$, i.e.,
    \[
    \mathrm{win}_\ell(D_i) = \{D_i, D_i+1, \dots, D_i+\ell-1\} \pmod n.
    \]
    Thus $\{D_i\}$ is uniformly distributed over all $n$ possible window starts. Suppose the continuous arcs $\{[C_j,C_j+a)\pmod 1\}$, with $a=\ell/n$ already, cover the entire circle.  
    Then for every index $i\in\{0,1,\dots,n-1\}$, the point $x_i=i/n$ must lie inside some arc $[C_r,C_r+a)$.  
    Writing $D_r=\lfloor n C_r \rfloor$, this immediately implies $i\in\mathrm{win}_\ell(D_r)$.  
    Hence full coverage in the continuous model implies full coverage in the cyclic-discrete model. 
   
    \label{cyclic vs continous proof}
    \end{IEEEproof}
    
    This upper bound helps concluding the asymptotic behavior of Problem~\ref{cyclic windows problem}.
    
    \begin{corollary}\label{cyclic windows asymptotic}
    For $\ell=o(n)$, it holds that 
    \[\mathbb{E}\!\big[T_{n,\ell}^\circlearrowright\big]\approx\tfrac{n}{\ell}\log {\big(\tfrac{n}{\ell}\big)}.
    \]
    \end{corollary}
    \begin{IEEEproof}
    As mentioned in Section \ref{sec:defs}, it was shown that \(\E[\mathcal{T}_a]\approx \frac{1}{a}\,\log\!\Big(\frac{1}{a}\Big)
    \text{, as } a\rightarrow 0.\)
    According to this and Lemma \ref{lem:cyc-vs-cont}, choosing $a=\frac{\ell}{n}$, concludes that for $\ell=o(n)$, \(\mathbb{E}\!\big[T_{n,\ell}^\circlearrowright\big]\lesssim \frac{n}{\ell}\,\log\!\Big(\frac{n}{\ell}\Big).\) Let $m=\lfloor \frac{n}{\ell}\rfloor$ and take representatives \(S=\{0,\ell,2\ell,\dots,(m-1)\ell\}.\) Any length-$\ell$ cyclic window covers only one element of $S$, and for each
    $i\in S$ the hit probability per draw is $\frac{\ell}{n}$. Thus the problem of covering $S$ reduces to the coupon collector's problem (\tight\cite{demoivre1713, laplace1812, bernstein1945, feller1950, erdos1961}) with \(\lfloor\frac{n}{\ell}\rfloor\) coupons. Therefore, \(\mathbb{E}\!\big[T_{n,\ell}^\circlearrowright\big]\ \ge \lfloor\frac{n}{\ell}\rfloor\,H_{\lfloor \frac{n}{\ell}\rfloor}
    \ \gtrsim\ \frac{n}{\ell}\,\log{\big( \frac{n}{\ell} \big) }.
    \)
    \end{IEEEproof}
    Corollary~\ref{cyclic windows asymptotic} shows the intuitive observation that as $n$ grows and $\ell=o(n)$, the discrete cyclic model shares the same leading asymptotic as the continuous circle covering problem. Now similarly to Section \ref{sec: non cyclic problem}, the general asymptotic regime where $\ell=\Omega (n)$ is examined, completing the asymptotic analysis of $\E[T_{n,\ell}^\circlearrowright]$ for any $\ell\le n$.

   \begin{corollary}\label{cor:cyclic-constant-factor}
    Fix a constant $c\ge2$. For $\frac{n}{c}\le \ell\ <\frac{n}{c-1}$, it holds that
    \[
    \mathbb{E}\!\big[T_{n,\ell}^{\circlearrowright}\big]=O(1).
    \]
    Furthermore, it holds that
    \[
    \mathbb{E}\!\big[T_{n,\ell}^{\circlearrowright}\big]
    \ \le\ \mathbb{E}\!\big[\mathcal{T}_{\ell/n}\big]
    \ \le\ \mathbb{E}\!\big[\mathcal{T}_{1/c}\big]
    \le1+c^2
    .\]
    \end{corollary}
    
    \begin{IEEEproof}
    By Lemma~\ref{lem:cyc-vs-cont} (Section \ref{cyclic vs continous}), $\mathbb{E}[T_{n,\ell}^{\circlearrowright}] \le \mathbb{E}[\mathcal{T}_a]$ with $a=\frac{\ell}{n}$. If $\ell\ge \frac{n}{c}$ then $a\ge \frac{1}{c}$, and since covering gets easier as $a$ increases, $\mathbb{E}[\mathcal{T}_a]\le \mathbb{E}[\mathcal{T}_{1/c}]$. Moreover, for $t$ samples of arcs, full coverage not occurs if and only if for every arc, the next one is at distance more than $a$ from it (which happens with probability $1-a$). By a union bound, the probability that this happens for all $t$ samples is at most $t(1 - a)^{t - 1}$. Hence,
\begin{align*}    
\mathbb{E}[T_a]
=& \sum_{t=0}^{\infty} \Pr[T_a > t]
= 1 + \sum_{t=1}^{\infty} \Pr[T_a > t]
\;\\\le&\;
1 + \sum_{t=1}^{\infty} t(1 - a)^{t - 1}
= 1 + \frac{1}{a^2}
\;\le\;
1 + c^2,
\end{align*}
showing that the bound depends purely on the ratio between $n$ and $\ell$.
    \end{IEEEproof}

   Going back to Problem~\ref{non cyclic windows problem}, in order to analyze the asymptotic behavior of the expectation in this problem, we bound it from both sides, which yields the next result.
    \begin{theorem}\label{asymptotics for non cyclic theorem}
    For $\ell=\omega(\log (n))$, it holds that
      \[
      \mathbb{E}[T_{n,\ell}]
      \;=\;\frac{3}{2}\,(n-\ell+1)\,\big(1+o(1)\big),
      \]
      and for $\ell=o(\log n)$, it holds that
      \[
      \mathbb{E}[T_{n,\ell}]
      \;=\;\frac{n}{\ell}\,\log\!\Big(\frac{n}{\ell}\Big)\,\big(1+o(1)\big).
      \]
      \end{theorem}
    
     \begin{IEEEproof} The coverage expectation is at most the sum of expectations of covering the extremes $[0,\ell-1]\cup[n-\ell,n-1]$, and the complimentary middle regime $[\ell,\,n-\ell-1]$ seperately. Roughly speaking, \(\mathbb{E}[T_{n,\ell}] \le \mathbb{E}[\text{cover extremes}] + \mathbb{E}[\text{cover middle}].\) As shown in Theorem~\ref{thm:large-windows}, \(
    \mathbb{E}[\text{cover extremes}] \;=\; \tfrac{3}{2}\,(n-\ell+1).\) Covering only the middle is no harder than covering a full cyclic instance with $n-\ell+1$ positions and window length $\ell$, with a probability for each position to be covered that is identical in both cases, being $\tfrac{\ell}{n-\ell+1}$. Therefore, \(\mathbb{E}[\text{cover middle}] \;\le\; \mathbb{E}\!\big[{T_{n-\ell+1,\ell}^{\circlearrowright}}\big]
    \;\le\;\E\big[\mathcal{T}_a\big]  \text{, with } a=\frac{\ell}{n-\ell+1}\), where the last inequality follows from Lemma~\ref{lem:cyc-vs-cont}. In addition, as mentioned in Section~\ref{sec:defs}, from \cite{flatto1962,after flatto - Steutel1967}, $\E[\mathcal{T}_{\frac{\ell}{n-\ell+1}}] \approx\frac{n-\ell+1}{\ell}\log(\frac{n-\ell+1}{\ell})\approx\frac{n}{\ell}\log(\frac{n}{\ell})$. Hence, the coverage time satisfies
    \(\mathbb{E}[T_{n,\ell}] \lesssim \tfrac{3}{2}\,(n-\ell+1) +\frac{n}{\ell}\log(\frac{n}{\ell})\).
    
    Additionally, the coverage time is at least as big as the coverage time to cover the extremes $\frac{3}{2}(n-\ell+1)$, and it is at least as big as the time to cover the representative points set \(S=\{\ell-1,2\ell-1,3\ell-1,\dots\, \lfloor \frac{n-\ell+1}{\ell}\rfloor\ell-1 \}\). Each window covers only one representative point from $S$, with probability $\frac{\ell}{n-\ell+1}$, therefore, The problem reduces to the coupon collector's problem (\tight\cite{demoivre1713, laplace1812, bernstein1945, feller1950, erdos1961}) with $\lfloor \frac{n-\ell+1}{\ell}\rfloor$ coupons. Hence, \(\mathbb{E}[T_{n,\ell}]\ \ge \lfloor\tfrac{n-\ell+1}{\ell}\rfloor H_{\lfloor\tfrac{n-\ell+1}{\ell}\rfloor }
    .\) This gives the explicit bounds,
    \(\label{eq:linear-upper}
    \max\big\{ \tfrac{3}{2}(n-\ell+1), \lfloor\tfrac{n-\ell+1}{\ell}\rfloor H_{\lfloor\frac{n-\ell+1}{\ell}\rfloor}\big\}
    \le
    \mathbb{E}[T_{n,\ell}].\) Therefore, by a simple analysis we get that for $\ell=\omega (\log (n)) $ the linear term dominates, and for $\ell=o(\log (n))$ the coupon collector term dominates, which makes the expectation asymptotically the same as in the cyclic case.
     \end{IEEEproof}

     \begin{observation}
    From Theorem~\ref{asymptotics for non cyclic theorem}, we observe that $\log n$ acts as a threshold on the expectation in the following way. Below it, the task is essentially  ‘‘find all middle positions’’, mirroring the cyclic behavior. Above it, the problem becomes essentially retrieving the two extremes.
    \end{observation}

    \section{The Batch Sampling Model and \\Coverage Monotonicity}\label{subsets and monotonicity section}    Problems~\ref{non cyclic windows problem},~\ref{cyclic windows problem} and~\ref{batch sampling problem} ask for the time to cover \(n\) items. Each model can be represented as hypergraph on \([n]\) as mentioned and described in detail in Section \ref{anina generalization}, with the collectable subsets of $[n]$ being the hyperedges of the corresponding Problem's hypergraph. In Problems \ref{cyclic windows problem} and \ref{batch sampling problem}, each item has the same degree (uniformity), each hyperedge has size $\ell$ (regularity) and the set of hyperedges is non-empty. We call any hypergraph on $[n]$ that satisfies those three conditions a \emph{uniform $\ell$-regular hypergraph on $[n]$}, or in short, a \emph{uniform $\ell$-regular model}. Denote the class of all uniform $\ell$-regular hypergraphs on $[n]$ as $\mathcal{U}_{n,\ell}$. For example, $\mathcal{U}_{n,1}$ consists of only one hypergraph denoted by $\mathcal{H}_{n,1}$ which is the hypergraph corresponding to the classical coupon collector's problem, having $\E[T_{\mathcal{H}_{n,1}}]=nH_n$. Note that for any $\mathcal{H}\in\mathcal{U}_{n,\ell}$, the set of hyperedges is a subset of all possible hyperedges admissable in Problem \ref{batch sampling problem} denoted as $\binom{[n]}{\ell}$. Denote the hypergraph of Problem~\ref{batch sampling problem} as $\mathcal{H}_{\binom{[n]}{\ell}}$. Note that $T_{\mathcal{H}_{\binom{[n]}{\ell}}}$ is the same as the random variable in Problem \ref{batch sampling problem}, $T_{\binom{[n]}{\ell}}$, and will be referred to as such from now on.
    
    In this section a natural question is suggested. Given $\mathcal{H}_1,\mathcal{H}_2\in\mathcal{U}_{n,\ell}$, with the corresponding sets of hyperedges $\mathcal{F}_1,\mathcal{F}_2$,
    does it holds that if $\mathcal{F}_1\subseteq \mathcal{F}_2$, or even if $|\mathcal{F}_1|\le|\mathcal{F}_2|$, then $\E[T_{\mathcal{H}_1}]\le\E[T_{\mathcal{H}_2}]$? If so, $\mathcal{H}_{\binom{[n]}{\ell}}$ with $T_{\binom{[n]}{\ell}}$ would have the highest expectation among $\mathcal{U}_{n,\ell}$,
    and any model with $\frac{n}{\ell}$ disjoint subsets would admit the lowest expectation. That is, for any $\mathcal{H}\in\mathcal{U}_{n,\ell}$, 
    $\lfloor\frac{n}{\ell}\rfloor H_{\lfloor\frac{n}{\ell}\rfloor}\le\E[T_{\mathcal{H}}] \le \E[T_{\binom{[n]}{\ell}}]$, providing universal upper and lower bounds on the expectation for the entire class.

    As mentioned in Section \ref{sec:defs},
    combinatorial solutions for Problem \ref{batch sampling problem} and the generalization, given
    \(A\subseteq[n]\) with \(|A|=m\), and \(Z_k(A)\) being the number of draws
    needed to see at least \(k\le m\) distinct elements of \(A\), exact formulas were obtained. While exact, those expressions are numerically delicate for large
    parameters due to large binomial coefficients and cancellations, and is hard to compute in a regular calculator. The next results gives tight lower and upper bounds on the expectation of this problem, and furthermore for any generalization, as explained below.
    
    While finalizing this paper, it was noticed that a very recent and concurrent work (\tight\cite{berend2025}) independently presented a similar proof to the one presented here. We wish to acknowledge their contribution and give due credit. Nevertheless, the proof is included here, as we believe it serves a broader purpose, demonstrating that the same reasoning naturally extends to generalized versions of the coupon collector's problem, including problems such as the double dixie cup and other variants. We also think and hope that the intuition in this proof can be applied in other coverage problems.   
    
    \begin{theorem}\label{eq:batch-sandwich}
    For any \(1 \le\ell<n\) and \(1\le k\le m\le n\) and $A\subseteq [n]$ with $|A| = m$, it holds that
    \[
    \frac{H_m-H_{m-k}}{\,H_n-H_{n-\ell}\,}
    \;\le\;
    \mathbb{E}[Z_k(A)]
    \;\le\;
    \frac{H_m-H_{m-k}-\frac{1}{n}}{\,H_n-H_{n-\ell}\,}+1 .
    \]
    \end{theorem}
    \begin{IEEEproof}\label{subset bounds proof}
    Fix $A\subseteq[n]$ with $|A|=m\le n$ and $k\le m$. Let $Y$ be the stopping time of the classical coupon collector's problem with single coupon per draw, when $k$ items of $A$ were obtained. Since each new coupon from $A$ is obtained with probability $\frac{m-j}{n}$, after $j$ distinct ones have already been seen, it holds that
    \[
\mathbb{E}[Y]
  = \sum_{j=0}^{k-1} \frac{n}{m-j}
  = n\bigl(H_m - H_{m-k}\bigr).
\]
 Let
    $X:=Z_k(A)$ be the number of $\ell$-subset draws until the same coverage.
    From a single-coupon run $i_1,i_2,\dots,i_Y$, form consecutive \emph{blocks} by
    cutting as soon as the block contains $\ell$ distinct coupon labels (within the block). Then, it is possible to partition $i_1,i_2,\dots,i_Y$ as follows.
    \begin{align*}
    i_1,\ldots,i_Y
    = \ &
    \underbrace{i_1,\ldots,i_{\ell+Z_1}}_{\ell+Z_1}\;,\;
    \underbrace{i_{\ell+Z_1+1},\ldots,i_{2\ell+Z_1+Z_2}}_{\ell+Z_2}\;,\; \ldots\;,\\[2pt]
    \quad&
    \underbrace{i_{(X-2)\ell+\sum_{j=1}^{X-2} Z_j + 1},\ldots,
               i_{(X-1)\ell+\sum_{j=1}^{X-1} Z_j}}_{\ell+Z_{X-1}}\;,\;
               \\&
    \underbrace{i_{(X-1)\ell+\sum_{j=1}^{X-1} Z_j + 1},\ldots,i_Y}_{\text{partial block}} \, .
    \end{align*}
    The $j$-th block then has length
    \(
    L_j=\ell+Z_j\), where 
    \( Z_j\ge 0\) (wasted repeats inside the block).
    The expected number of draws until $\ell$ distinct items were drawn is 
    \[\E[L_j]=\sum_{k=1}^{\ell}\tfrac{n}{n-k+1}=n(H_n-H_{n-\ell}).\]
    Since $Z_j=L_j-\ell$, we get
    \[
    \E[Z_j]=\E[L_j]-\ell
    = n\!\left(H_n - H_{n-\ell}\right) - \ell.
    \]
    To derive the bounds, consider the following. After $X-1$ blocks, the coverage process was almost finished, and the last draws until full coverage does not necessarily make a full block. Let $D\ge 1$ be the extra draws needed to reach full coverage after $X-1$ blocks were drawn (by definition, $X-1$ draws had not yet covered all items and at most one block is required to complete coverage). Then
      \[
    Y \;=\; \sum_{i=1}^{X-1} (\ell+Z_i) \;+\; D.
    \]
    Taking expectations gives
    \[
    n\big(H_m -H_{m-k}\big) - \E[D] \;=\; \ell\,\E[X-1] \;+\; \E[X-1]\E[Z],
    \]
    which yields
    \begin{equation}
    \label{eq:Z_batch-bounds}
    \E[X] = 1 + \frac{H_m\tight -\tight H_{m-k} \tight-\tight \frac{\E[D]}{n}}{H_n - H_{n-\ell}}.  
    \end{equation}
    By the definition of $D$, and since in the last $D$ draws the number of different collected items is at most $\ell$, it holds that  $1\le\E[D]\le\E[L_j]$. Therefore, by substituting into Equation \ref{eq:Z_batch-bounds} we get the final bounds
    \[
     \frac{H_m\tight -\tight H_{m-k}}{H_n - H_{n-\ell}}\le
     \E[X]
    \le 1\tight +\tight \frac{H_m\tight -\tight H_{m-k}\tight-\tight \frac{1}{n}}{H_n - H_{n-\ell}}.
    \]
    \end{IEEEproof}  
Using Theorem~\ref{eq:batch-sandwich}, tight lower and upper bounds for Problem~\ref{batch sampling problem} are obtained. The special case of $A=[n]$, and $k=n$, yields that for any \(1\le \ell\le n\),
    \[
    \frac{H_n}{\,H_n-H_{n-\ell}\,}
    \;\le\;
    \mathbb{E}[T_{\binom{[n]}{\ell}}]
    \;\le\;
    \frac{H_{n-1}}{\,H_n-H_{n-\ell}\,}+1,
    \]
which coincides with the result of~\cite{berend2025}.

Next, more powerful statement is presented, explaining how Theorem \ref{eq:batch-sandwich} can be extended. We define the term \emph{coverage task} as getting a collection that satisfies some requirement. The coupons collector's problem has a coverage task that requires to obtain each coupon at least once. Another coverage task can be to obtain any coupon twice, as in the double dixie cups problem (\tight\cite{double-dixie1960}), or maybe to either get each one twice, or to get half of the coupons three time. Let $\mathcal{C}$ be a coverage task. We can consider both $\mathcal{H}_{n,1}$ and $\mathcal H_{\binom{[n]}{\ell}}$ to be the hypergraphs from which hyperedges are being sampled, and we can ask about the relation of expected number of steps to complete $\mathcal{C}$ in both models. Denote $Q ^{\mathcal{C}}_{{n,1}}$ as the number of hyperedges drawn from $\mathcal H_{n,1}$ until completing $\mathcal{C}$, and denote $Q ^{\mathcal{C}}_{{n,\ell}}$ as the number of draws until completing $\mathcal{C}$ drawing from $\mathcal{H}_{\binom{[n]}{\ell}}$.
    \begin{theorem}
    \label{thm:subset-normalization}
    Let $\mathcal{C}$ be any coverage task. It holds that
    \[
    \frac{\mathbb{E}[Q^\mathcal{C}_{n,1}]}{\,n\!\left(H_n - H_{n-\ell}\right)} 
    \;\le\;
    \mathbb{E}[Q^{\mathcal{C}}_{n,\ell}]
    \;\le\;
    \frac{\mathbb{E}[Q^\mathcal{C}_{n,1}]-\frac{1}{n}}{\,n\!\left(H_n - H_{n-\ell}\right)} + 1,
    \]
    \end{theorem}

\begin{IEEEproof}
The claim follows directly from the argument in the proof of Theorem~\ref{eq:batch-sandwich}, substituting \(X = Q_{n,\ell}^{\mathcal{C}}\) and \(Y=Q_{n,1}^{\mathcal{C}}\).
\end{IEEEproof}
 This shows that any coverage task inherits the same normalization factor that governs the transition from using the single-coupon draws of $\mathcal{H}_{n,1}$ to the $\ell$-subset draws of $\mathcal{H}_{\binom{[n]}{\ell}}$.

One general framework that encompasses a broad family of coupon collector's problem coverage tasks can be formulated as follows. Let
\[
    \bfv = (v_1, v_2, \ldots, v_n) \in \mathbb{N}^n
\]
be a demand vector, where $v_i$ specifies the required number of times item~$i$
must be collected.
We let $Q_{n,1}^\bfv$ denote the number of draws $\mathcal{H}_{n,1}$'s hyperedges required for obtaining each $i\in[n]$ at least $v_i$ times, and $Q_{n,\ell}^\bfv$ as number the of $\mathcal H_{\binom{[n]}{\ell}}$'s hyperedges needed to complete the same task.

The double dixie cup problem mentioned above is a classical special case, obtained by setting a constant $m$, and taking 
$\bfv = (m,m,\ldots,m)$.
In this setting (\tight\cite{double-dixie1960}) it is known that
\[
    \mathbb{E}[Q_{n,1}^\bfv]
    = n \Bigl(\log n + (m-1)\log \log n + \gamma + o(1)\Bigr).
\]
By Theorem~\ref{thm:subset-normalization}, it is obtained that
\[
    \mathbb{E}[Q_{n,\ell}^\bfv]
    \ge \frac{n \Bigl(\log n + (m-1)\log \log n + \gamma + o(1)\Bigr)}
    {n(H_n-H_{n-\ell})}
,\]
and
\[
    \mathbb{E}[Q_{n,\ell}^\bfv]
    \le \frac{n \Bigl(\log n + (m-1)\log \log n + \gamma + o(1)\Bigr)-\frac{1}{n}}
    {n(H_n-H_{n-\ell})}+1.
\]
Note that the gap between the lower and upper bounds is at most 1, making the bounds tight.

Addressing the monotonicity question, Problem~\ref{cyclic windows problem} is the
    parent of a class of problems we call \emph{$\Delta$-d window start set}. For integers $n,\ell,d$ with
    $\ell\mid n$ and $d\mid\ell$, denote $\mathcal{U}^{\circlearrowright}_{n,\ell}\subseteq \mathcal{U}_{n,\ell}$ to be the class of hypergraphs with hyperedges that are cyclic windows. Denote the hypergraph of Problem \ref{cyclic windows problem} as $\mathcal{H}^\circlearrowright_{n,\ell}$, and its set of hypersdges $\mathcal{F}^\circlearrowright_{n,\ell}$. Note that for each $\mathcal{H}_d\in\mathcal{U}^\circlearrowright_{n,\ell}$ with hypersdges set $\mathcal{F}_{d}\subseteq \mathcal{F}^\circlearrowright_{n,\ell}$ which is referred to as \emph{$\Delta$-d start set}, and the windows start positions are at distance $d$ apart. Note that $\mathcal{H}_d\in\mathcal{U}_{n,\ell}$ is equivalent to $\mathcal{H}^\circlearrowright_{\frac{n}{d},\frac{\ell}{d}}\in\mathcal{U}_{\frac{n}{d},\frac{\ell}{d}}$, which is the hypergraph of Problem \ref{cyclic windows problem} setting $n' = n/d$ and $\ell' = \ell/d$ (since there are $\frac{n}{d}$ $d$-lengthed blocks, and on each block start, a window starts, containing $\frac{\ell}{d}$ blocks). As a consequence, proving the monotonicity for the \emph{$\Delta$-d} window class $\mathcal{U^\circlearrowright}_{n,\ell}$ is equivalent to proving the next lemma.\footnote{one may ask why wasn't the expression in Theorem \ref{thm:general-cyclic} used in the proof. It is due to the complex nature of the inclusion-exclusion terms that makes it hard to compare.}
    \begin{lemma}
    \label{lem:d stride-monotone}
    For any $n,\ell,d$ with $d\mid \ell \mid n$, it holds that
    \[
    \mathbb{E}\!\left[T^{\circlearrowright}_{\,\frac{n}{d},\;\frac{\ell}{d}}\right]
    \;\le\;
    \mathbb{E}\!\left[T^{\circlearrowright}_{\,n,\;\ell}\right].
    \]
    Equivalently, within the \emph{$\Delta$}-d family with fixed $n$ and $\ell$, making starts more discrete decreases the expected coverage time.
    \end{lemma}
    \begin{IEEEproof}
    The proof is a coupling argument, very similar to the proof of Lemma~\ref{lem:cyc-vs-cont}.
    Fix $d\mid \ell\mid n$ and run the cyclic $(n,\ell)$ model with starts uniformly on $\{0,1,\dots,n-1\}$ and windows $\mathrm{win}^{\circlearrowright}_\ell(i)$.
    Define
    \[
    \phi(i)=d\Big\lfloor \tfrac{i}{d}\Big\rfloor \in \{0,d,2d,\dots,n-d\},
    \]
    which is uniform over the \emph{$\Delta$-d} start set (each value has $d$ preimages),
    so using $\phi$ gives a uniform \emph{$\Delta$-d} process (which justifies the coupling). In other words, the mapping shifts windows to start at multiples of $d$ (block starts).
    Let $\mathcal{R}=\{w_1,\dots,w_{|\mathcal R|}\}$ be the windows drawn up to
    $T^{\circlearrowright}_{n,\ell}$ (they cover all $n$ positions).
    Consider any $d$-block $[jd,(j+1)d-1]$.
    If it was covered by a window starting at $jd$, it remains covered after applying $\phi$.
    If not, some window in $\mathcal R$ starts inside this block. So applying $\phi$
    moves the window start to position $jd$, covering the whole block.
    Thus all blocks are covered by the $\phi$-mapped sequence. Therefore, coverage in the cyclic windows model implies coverage in the \emph{$\Delta$-d} model.
    \end{IEEEproof}
    \begin{remark}
    The same coupling argument extends naturally to higher dimensions.  
    For example, consider a $2$-dimensional torus of size $n_1\times n_2$, where each sample is a rectangular patch of $\ell_1\times \ell_2$.  
    If $d_1\mid \ell_1\mid n_1$ and $d_2\mid \ell_2\mid n_2$, we can partition the torus into $d_1\times d_2$ blocks and define the mapping
    \[
    \phi(i,j)=\big(d_1\lfloor i/d_1\rfloor,\; d_2\lfloor j/d_2\rfloor\big),
    \]
    which is uniform over the \emph{$\Delta$}-$(d_1,d_2)$ start set.  
    The same block-covering argument shows that coverage in the full model implies coverage in the \emph{$\Delta$}-d model, yielding the same monotonicity property.

    This monotonicity extends inductively to any dimension.  
    Moreover, there is always a corresponding continuous version obtained by letting $n_i\to\infty$ while fixing the ratios $\ell_i/n_i$, which dominates all discrete versions in expectation (by the same argument proving Lemma~\ref{lem:cyc-vs-cont}).  
    \end{remark}

    \begin{remark}
    \label{cyclic-torus-lower-bound}
        For Problem \ref{cyclic windows problem}, a lower bound of $\lfloor\frac{n}{\ell}\rfloor H_{\lfloor\frac{n}{\ell}\rfloor}$ had been derived in the proof of Lemma \ref{cyclic windows asymptotic}. In the special case where $\ell\mid n$ this mathces with a general lower bound that is true for all models, established in \cite{chang-and-fang} for $n=2$, and as stated in \cite{berend2025}, can be extended to any $\ell \mid n$.

    \end{remark}
    These monotonicity results suggest three universal principles which we believe are true.
    \begin{conjecture}
    \label{con:monotonicity}
         Given $\mathcal{H}_1,\mathcal{H}_2\in\mathcal{U}_{n,\ell}$, with the corresponding sets of hyperedges $\mathcal{F}_1,\mathcal{F}_2$ such that $\mathcal{F}_1\subseteq \mathcal{F}_2$, it holds that
        \[        \E[T_{\mathcal{H}_1}]\le\E[T_{\mathcal{H}_2}].
        \]\end{conjecture} 
    \begin{conjecture}
    \label{con:batch-universal-bound}
        Given $\mathcal{H}\in\mathcal{U}_{n,\ell}$, it holds that
        \[
        \E[T_\mathcal{H}]\le\E[T_{\binom{[n]}{\ell}}].
        \]
    \end{conjecture}
        \begin{conjecture}
    \label{con:batch-universal-lower-bound}
        Given $\mathcal{H}\in\mathcal{U}_{n,\ell}$, it holds that
        \[
        \lfloor\tfrac{n}{\ell}\rfloor H_{\lfloor \frac{n}{\ell}\rfloor} \le\E[T_\mathcal{H}].
        \]
    \end{conjecture}
    The conjectures suggest a universal upper bound on the uniform $\ell$-regular models. Furthermore, a conjecture by~\cite{ahu-schilling}, later proved by~\cite{berend2025}, stated that the uniform distribution, corresponding to the batch coverage model in our terminology, does not achieve the minimal expected time. Our conjecture broadens the horizon of this idea and claims that not only is this model not optimal, but may even be the worst within the entire class $\mathcal{U}_{n,\ell}$.
    
    Although the conjectures, and in particular Conjecture~\ref{con:batch-universal-bound} remains currently unproved, a slightly weaker bound can already be established.
    The following theorem states this bound explicitly.
             
    \begin{theorem}
    \label{thm:universal-upper}
    Given $\mathcal{H}\in\mathcal{U}_{n,\ell}$, it holds that
    \[
    \mathbb{E}[T_\mathcal{H}]\;\le\;\frac{n}{\ell}\,\big(\log n + 1\big).
    \]
    \end{theorem}
    
    \begin{IEEEproof}
    Let $U_t$ be the number of items not yet seen after $t$ draws. Due to uniformity, the probability for a position to not be sampled after $t$ rounds is \(\big(1-\frac{\ell}{n}\big)^t\).
    Thus, $\mathbb{E}[U_t]=n(1-\ell/n)^t$. Using the tail-sum formula and Markov’s inequality,
    \(
    \mathbb{E}[T_\mathcal{H}]
    =\sum_{r=1}^{\infty}\Pr[T_\mathcal{H}\ge r]
    =\sum_{r=1}^{\infty}\Pr[U_{r-1}\ge 1]
    \;\le\;\sum_{r=1}^{t_0} 1 \;+\;\sum_{r=t_0+1}^{\infty}\mathbb{E}[U_{r-1}],
    \)
    where we choose $t_0:=\left\lceil \tfrac{n\log n}{\ell}\right\rceil$. Then
    \(
    \sum_{r=t_0+1}^{\infty}\mathbb{E}[U_{r-1}]
    = n\sum_{r=t_0}^{\infty}(1-\ell/n)^r
    = n\cdot\frac{(1-\ell/n)^{t_0}}{1-(1-\ell/n)}
    = \frac{n^2}{\ell}\,(1-\ell/n)^{t_0}.
    \)
    Using $(1-\ell/n)^{t_0}\le e^{-(\ell/n)t_0}\le e^{-(\ell/n)\cdot (n\log n/\ell)}=e^{-\log n}=1/n$, we get
    \[
    \sum_{r=t_0+1}^{\infty}\mathbb{E}[U_{r-1}]\;\le\;\frac{n^2}{\ell}\cdot \frac{1}{n}\;=\;\frac{n}{\ell}.
    \]
    Therefore,
    \[
    \mathbb{E}[T_\mathcal{H}]\;\le\; t_0 + \frac{n}{\ell}
    \;\le\; \frac{n}{\ell}\log n + \frac{n}{\ell}
    \;=\; \frac{n}{\ell}\,(\log n + 1).
    \]
    \end{IEEEproof}

\section{Future Work}\label{sec:future-work}

A natural direction for future research is obtaining a more accurate asymptotic expression for the expectation in Problem~\ref{non cyclic windows problem}. 
Unlike the cyclic or continuous versions, Problem \ref{non cyclic windows problem} is harder to analyze, resulting in looser asymptotics which remain relatively crude.

Another important direction concerns the precision gap between the asymptotics of different models. For structured models such as the cyclic window (Problem~\ref{cyclic windows problem}), and its higher dimensional generalizations in $\mathcal{U}_{n,\ell}$, a lower bound of $\lfloor\frac{n}{\ell} \rfloor H_{\lfloor \frac{n}{\ell} \rfloor}$ was established as mentioned in Remark \ref{cyclic-torus-lower-bound}, and is believed to be universal for $\mathcal{U}_{n,\ell}$ as stated in Conjecture \ref{con:batch-universal-lower-bound}. A universal upper bound of $\frac{n}{\ell}(\log n + 1)$ was established for $\mathcal{U}_{n,\ell}$.

The two bounds share the same leading term \(\tfrac{n}{\ell}\log n\) when $\ell=o(n)$. On one hand, this is encouraging since it means that all such models are asymptotically solved. 
On the other hand, it highlights an important ambiguity: describing the expectation of a coverage problem as “asymptotically \(\tfrac{n}{\ell}\log n\)” is not very informative. 
Different models may yield the same asymptotic form, yet differ substantially in their non-leading terms. 

Although the lower and upper bounds behave asymptotically the same with respect to the dominant term (in particular, their ratio approaches one as $n$ increases), there is an of $O(\frac{n}{\ell}\log\ell)$ gap in their difference. Closing this gap or analyzing the expectation for each model separately is left for future work.
This suggests that further study of sub-leading terms and structural effects between models 
is essential to understand the fine behavior inside the regime where 
\(\tfrac{n}{\ell}\log(\frac{n}{\ell})\approx\tfrac{n}{\ell}\log n\).

Finally, proving or disproving either Conjecture \ref{con:monotonicity} or Conjecture \ref{con:batch-universal-bound}   could reveal deeper structural insights into coverage problems with uniform $\ell$-regular hypergraphs. Even a single counterexample or partial proof could illuminate the principles 
governing how geometry, overlap, and locality influence coverage time.

    \balance

\end{document}